\documentclass[12pt]{article}
\usepackage{amsmath}
\usepackage{hyperref}
\usepackage{amsfonts}
\usepackage{amssymb}
\usepackage{graphics}
\usepackage{mathdots}
\usepackage{wrapfig}
\usepackage{color}
\usepackage[pdftex]{graphicx}

\def\bee{\begin{equation}}
\def\eee{\end{equation}}

\def\Li{\rm {Li}}

\def\DHLhksqrt#1#2{\setbox0=\hbox{$#1\sqrt{#2\,}$}\dimen0=\ht0
\advance\dimen0-0.2\ht0
\setbox2=\hbox{\vrule height\ht0 depth -\dimen0}%
{\box0\lower0.4pt\box2}}
\begin{document}

\thispagestyle{empty}
\bigskip
\centerline{    }
\centerline{\Large\bf Will a physicist  prove the Riemann Hypothesis?}
\bigskip\bigskip
\centerline{\large\sl Marek Wolf}
\begin{center}
Cardinal  Stefan  Wyszynski  University, Faculty  of Mathematics and Natural Sciences. \\  
ul. W{\'o}ycickiego 1/3,   PL-01-938 Warsaw,   Poland, e-mail:  m.wolf@uksw.edu.pl
\end{center}

\bigskip\bigskip
\begin{abstract}
In the first part we present the number theoretical properties of the Riemann zeta function and formulate
the Riemann Hypothesis.  In the second part we  review some physical problems related to this hypothesis:
the  Polya--Hilbert conjecture, the links with Random Matrix Theory, relation with the Lee--Yang theorem on the zeros of
the partition function,   random walks, billiards etc.
\end{abstract}

\bigskip\bigskip


\bigskip\bigskip
\bigskip\bigskip

\bibliographystyle{ieeetr} 

\noindent\hspace*{4cm} ``... upon looking at prime numbers  one has the feeling of being  \\
\hspace*{4cm} ~~~in the presence of one of the inexplicable  secrets of creation.''\\
{\small \hspace*{9cm}Don Zagier  in \cite[p.8, left column]{Zagier1977}}

\section{Introduction}
\label{wstep}

There are many links between mathematics and physics. Many branches of mathematics arose from the need to formalize
and clarify the calculations carried out by physicists, e.g.  Hilbert spaces, distribution theory,  differential geometry etc.
In   this  article we are going to describe  the opposite situation when the famous open mathematical problem
can be perhaps solved by physical methods. We mean the Riemann Hypothesis (RH), the over 160 years old problem
which solution is  of central importance in many branches  of mathematics:  there are probably  thousands of theorems
beginning with: ``Assume that the Riemann Hypothesis is true, then ...''.
The RH appeared on the Hilbert's famous list  of problems for the XX centaury  as the first part of the eighth  problem  \cite{Hilbert-1902}
(second part  concerned the  Goldbach's conjecture; recently  H.~A. Helfgott   \cite{Helfgott-2015} have solved so called ternary
case of Goldbach conjecture).
In the  year  2000 RH appeared on the list of the Clay Mathematics Institute problems for the third millennium, this time with 1,000,000
US dollars reward for the solution.

After the announcement of the prize by Clay  Mathematics Institute for solving RH there  has been a rash of popular books
devoted to this problem:  \cite{Derbyshire}, \cite{Sabbagh2002}, \cite{Sabbagh-zeros}, \cite{Rockmore2005}, \cite{Sautoy2003}.
The classical monographs on RH are:  \cite{Titchmarsh}, \cite{Edwards},  \cite{Ivic1985}, \cite{Karatsuba-Voronin},
while \cite{Borwein_RH} is a collection of original papers devoted to RH preceded by extensive introduction to the subject.
We also strongly  recommend the web site {\it Number Theory and Physics}  at the address
\href{http://empslocal.ex.ac.uk/people/staff/mrwatkin/zeta/physics.htm}{http://empslocal.ex.ac.uk/people/staff/mrwatkin/zeta/physics.htm}
containing a lot information about links between number theory (in particular about RH) and physics.

In 2011  there appeared the paper  ``Physics of the Riemann hypothesis''  written by D. Schumayer  and D. A.W. Hutchinson
\cite{Physics-and-RH-RevModPhys}.  Here we aim to provide  complementary description of the problem which can serve as a
starting point for the interested reader.    

This review consists of seven Sections and the Concluding Remarks. In the next Section  we present the historical path leading
to the formulation of the RH. Next we briefly  discuss possible ways of proving the RH. Next two Sections concern
connections between  RH and quantum mechanics and statistical mechanics. In Sect.\ref{RW} a few other links between physical
problems  and RH  are presented.  In the last Section fractal structure of the Riemann $\zeta(s)$ function is discussed.
Because we intend this  article  to be a guide we enclose rather exhaustive  bibliography containing over 100 references,
a lot of these papers can be  downloaded freely from author's web pages.  Let us begin the story.

\section{The short history of the  Prime Number Theorem}
\label{historia}

There are infinitely many prime numbers $2, 3, 5, 7, 11, \ldots, p_n, \ldots$  and the first proof of this fact was given
by Euclid in his {\it Elements} around 330 years b.C. His proof was by contradiction: Assume there is a finite set of primes
$\mathcal{P}=\{2, 3, 5, \ldots, p_n\}$. Form the number $2\times 3 \times 5\ldots \times p_n+1$,
then this number divided by  primes from $\mathcal{P}$ gives the remainder 1, thus it has to be a new prime or it has to factorize
into primes not contained in the set $\mathcal{P}$, hence there must be more primes than $n$. For example if $\mathcal{P}=
\{2,3,5,7,11, 13\}$, then $2\cdot 3\cdot 5\cdot 7\cdot 11\cdot 13 + 1=59\cdot 509$ and $59, 509 \notin \mathcal{P}$.
The first direct proof of infinity of primes was presented by L. Euler around 1740 who has shown that the harmonic sum of
prime numbers $p_n$ diverges:
\[
\sum_{p_n<x}~\frac{1}{p_n}\sim \log \log(x).
\]
Next  the problem of  determining  the function $\pi(x)=\sum_{n} \Theta(x-p_n)$  ($\Theta(x)$ is the  Heaviside step function),
giving the number of primes up to a given
threshold $x$,  has arisen. It is one of the greatest surprises in the whole mathematics that such an erratic function
as $\pi(x)$ can be approximated by a simple expression. Namely  Carl Friedrich Gauss as a teenager (different sources put his age
between  fifteen years and seventeen years)  has made in the end of eighteen centaury  the conjecture that
$\pi(x)$ is very well approximated  by the logarithmic integral $\Li(x)$:
\bee
\pi(x)\sim {\Li}(x):=\int_2^x  \frac{du}{{\log(u)}}. 
\label{PNT}
\eee
The  symbol $f(x) \sim  g(x)$ means here that $\lim_{x\rightarrow \infty} f(x)/g(x)=1$.
Integration by parts gives the asymptotic expansion which should be cut at the term $n_0=\lfloor \log(x)\rfloor$,  after which
terms begin to increase:
\bee
{\rm Li}(x) = \frac{x}{\log(x)}  + \frac{x}{\log^2(x)} + \frac{2!x}{\log^3(x)} + \frac{3!x}{\log^4(x)} +\cdots .
\label{Li2}
\eee
There is a series giving $\Li(x)$ for  all $x>2$ and  quickly convergent which has $n!$ in denominator and $\log^n(x)$
in nominator instead of opposite order in (\ref{Li2})
(see \cite[Sect. 5.1]{abramowitz+stegun})
\begin{equation}
{\Li} (x) =  \gamma +\log \log(x) + \sum_{n=1}^{\infty} \frac{\log^{n}(x)}{n \cdot n!}\quad {\rm for} ~ x > 1 ~ ,
\label{Li-series}
\end{equation}
where $\gamma \approx 0.577216$ is the Euler-–Mascheroni constant  $\gamma = \lim_{n \rightarrow \infty}  \left( \sum_{k=1}^n \frac{1}{k} - \log(n) \right)$.

The way of proving \eqref{PNT} was outlined  by Bernhard Riemann in a  seminal 8-pages long  paper
published in 1859 \cite{Riemann-1859}. English translation is available at  \href{http://www.maths.tcd.ie/pub/HistMath/People/Riemann}{http://www.maths.tcd.ie/pub/HistMath/People/Riemann}
or at \href{http://www.claymath.org/sites/default/files/ezeta.pdf}{CMI web page}; it was also included as an appendix in
\cite{Edwards}.  The  handwritten by Riemann  manuscript was saved by his wife and is kept in the  Manuscript Department of the
Nieders{\"o}chsische Staats und Universit{\"a}tsbibliothek G{\"o}ttingen.  The scanned pages are available
at  http://www.claymath.org/sites/default/files/riemann1859.pdf.   In fact in this paper Riemann has given an {\it exact}
formula for $\pi(x)$.  The starting point of the Riemann's reasoning  was  the mysterious formula   discovered by Euler
linking the sum of  $\frac{1}{n^{ks}}$  with the product over all  primes $p$:
\bee
\zeta(s) := \sum_{n=1}^\infty \frac{1}{n^s} = \prod_{p=2}^\infty \frac{1}{\left(1-\frac{1}{p^s}\right)}, ~~~~s=\sigma+it, ~~~~\Re[s] = \sigma >1.
\label{gold-formula}
\eee
To see that this equality really holds one needs first to recognize in the terms $1/{\left(1-\frac{1}{p^s}\right)}$ the sums
of the  geometric  series  $\sum_{k=0}^\infty \frac{1}{p^s}$. The geometrical  series converges absolutely so the interchange of
summation and the product is justified.  Finally the  fundamental theorem of arithmetic stating the each positive integer $n > 1$
can be  represented in exactly one way  (up to the order of the factors)  as a product of prime powers:
\bee
    n = p_1^{\alpha_1}p_2^{\alpha_2} \cdots p_k^{\alpha_k} = \prod_{i=1}^{k}p_i^{\alpha_i}.
\label{fta}
\eee
has to be invoking  to obtain the
series on lhs of \eqref{gold-formula}.   Let us notice that on the rhs \eqref{gold-formula}  the product  cannot start from $p=1$ and it explains why the first prime is 2 and not 1 ---
physicists often think that 1 is a prime number (before  19--th century $1$  was indeed considered to be a prime).
Euler was the first who calculated the sums $\sum_{n=1}^\infty \frac{1}{n^2}=\pi^2/6,~
\sum_{n=1}^\infty \frac{1}{n^4}=\pi^4/90$ and in general $\zeta(2n)$. In fact Euler has considered the above formula only
for real exponents $s=x$ and it was Riemann who considered it as
a function of complex argument $s=\sigma+it$ and thus the function $\zeta(s)$  is called
the Riemann's zeta function.  In the context of RH instead of $z=x+iy$ for the complex variable the notation $s=\sigma+it$  is
traditionally used. The formula (\ref{gold-formula}) is valid only for $\Re[s] = \sigma >1$
and it follows from the product of non-zero terms on  r.h.s. of  \eqref{gold-formula} that $\zeta(s)\neq 0$ on the right
of the line $\Re[s]=1$.  Riemann has  generalized  $\zeta(s)$ to the
whole complex plane without $s=1$ where zeta is divergent as an usual harmonic series  --- the fact established in
14th century by Nicole Oresme. Riemann did it by analytical continuation  (for the  proof see  the original  Riemann's paper
or  e.g.  \cite[Sect. 1.4]{Edwards}:
\bee
\zeta(s)=\frac{\Gamma(-s)}{2\pi i}\int_{+\infty}^{+\infty} \frac{(-x)^s}{e^x-1} \frac{dx}{x},
\label{zeta-plane}
\eee
\noindent
\noindent where  $\int_{+\infty}^{+\infty}$ denotes the integral over the contour\\

\centerline{      }

\centerline{      }

\centerline{      }

\centerline{      }

\centerline{      }

\begin{picture}(0,0)(0,0)

\put(240,30){\vector(0,1){60}}
\put(180,60){\vector(1,0){120}}

\put(240,60){\oval(20,20)[l]}
\put(240,65){\oval(10,10)[rt]}
\put(240,55){\oval(10,10)[rb]}

\put(245,55){\vector( 1,0){45}}
\put(290,65){\vector(-1,0){45}}

\end{picture}
\vskip-1cm
\noindent Appearing in  \eqref{zeta-plane}  the  gamma function  $\Gamma(z)$  has many representations, we present
the Weierstrass product:
\bee
\Gamma(z) = \frac{e^{-\gamma z}}{z} \prod_{k=1}^\infty \left(1 + \frac{z}{k}\right)^{-1} e^{z/k}\,.
\label{Gamma}
\eee
From (\ref{Gamma}) it is seen that $\Gamma(z)$  is defined for all complex numbers  $z$, except $z = -n$ for integers $n > 0$,
where are the  simple poles of $\Gamma(z)$.
The most popular definition of gamma function given by the integral $\Gamma(z) = \int_0^\infty e^{-t} t^{z-1} dt $
is valid only for $\Re[z]>0$.

The integral (\ref{zeta-plane}) is well  defined on the whole complex plane without $s=1$,
where $\zeta(s)$ has the simple pole, and is equal to (\ref{gold-formula})
on the right  of the line $\Re[s]=1$. Recently many representations of the $\zeta(s)$ are known,
for review of the integral representations see \cite{Milgram-2013}.

The {\it exact} formula for $\pi(x)$ obtained by Riemann involves the function $J(x)$ defined  as
\bee
J(x)  =  \pi(x) + \frac{1}{2} \pi(x^{1/2}) + \frac{1}{3} \pi(x^{1/3}) + \cdots.
\label{J_od_x}
\eee
In words $J(x)$  increases by 1 at each prime number $p$,  by $1/2$ at each $x$ being the square of the prime, in general
it jumps by $1/n$ at argument equal to $n$-th power of some prime $p^n$. $J(x)$ is discontinuous at $x=p^n$ thus at such arguments
is defined as a halfway between its old value and its new value.  Let us mention that $J(x)=0$ for $0\leq x < 2$. Then  $\pi(x)$
 is given (via so called M{\"o}bius inversion formula)  by the series:
\bee
\pi (x) = \sum_{n\geq 1} \frac{\mu (n)}{n}  J(x^{1/n}),
\label{pi-od-j}
\eee
where the sum is in fact finite because it stops at such $N$ that $x^{1/N}>2>x^{1/(N+1)}$ and   $\mu(n)$
is the  M{\"o}bius function:
\bee
\mu(n) \,=\,
\left\{
\begin{array}{ll}
1 & \mbox {for}~~   n =1 \\
0 & \mbox {when {\it n} is divisible by the square of some prime {\it p}:} ~~p^2|n\\
(-1)^r & \mbox{ when }~ n=p_1 p_2 \ldots p_r
\end{array}
\right.
\label{mobius}
\eee
For example $ \mu(14) = 1, \mu(25)=0, \mu(30) = -1$. This definition of $\mu(n)$ stems from   the formula (\ref {gold-formula})
and  the Dirichlet series  for the reciprocal of the zeta function:
\bee
\frac{1}{\zeta(s)}=\prod_{p=2}^\infty \left(1-\frac{1}{p^s}\right)=\sum_{n=1}^\infty \frac{\mu(n)}{n^s}.
\label{inverse-zeta}
\eee
The above product over $p$ produces integers $n$ which in the factorization does not contain
square of a prime and those $n$ which factorizes into odd number of  primes contribute with sign $-1$ while
those $n$ which factorizes into even  number of  primes contribute with sign $+1$. We can  notice at this point
that the M{\"o}bius function has the physical interpretation:  namely in \cite{Spector-1990} it was shown that $\mu(n)$ can
be interpreted as the operator $(-1)^F$ giving the number of fermions in quantum field theory.  In this approach
the equality $\mu(n)=0$ for $n$   divisible by  a square of some prime is the manifestation of  the Pauli exclusion principle.

The crucial point of the Riemann's  reasoning was the alternative formula for $J(x)$ not {\it involving  primes at all}:
\bee
J(x)={\Li}(x)-\sum_{\rho}{\Li}(x^\rho),
\label{J-explicite}
\eee
where the sum runs over all zeros of $\zeta(s)$, i.e. over such $\rho$ that  $\zeta(\rho)=0$.   Let us stress that  the
above \eqref{J-explicite}  is an {\it equality}, which is remarkable because the left hand side is a step function,
thus somehow  at prime powers all of the zeros of zeta cooperate to deform smooth plot of the first term ${\Li}(x)$ into the
stair--like  graph with jumps.  Then the number of primes  up to $x$ is obtained by combining \eqref{pi-od-j} and
\eqref{J-explicite}
\bee
\pi (x) = \sum_{n=1}^N \frac{\mu (n)}{n} \left({\Li}(x^{1/n})-\sum_{\rho}{\Li}(x^{\rho/n})\right).
\label{pi-exact}
\eee
To be precise at arguments $x$ equal to prime numbers, when $\pi(x)$ is not continuous and jumps by 1, one has to define
lhs as $\lim_{\epsilon \rightarrow \infty} \{\pi(x-\epsilon)+\pi(x+\epsilon)\}$ (the same procedure was mentioned above
for the function $J(x)$).  There is an ambiguity when using definition of logarithmic integral \eqref{PNT} for ${\Li}(x^\rho)$
connected with  multivaluedness of logarithm of complex argument, in particular for complex
numbers  $z_1, z_2$ the equality $\log(z_1 z_2) = \log(z_1) + \log(z_2)$ does not hold (here are calculations providing the
counterexample:  $(-z)^2=z^2, ~\log((-z)^2)=\log(z^2), ~\log(-z)+\log(-z)=\log(z)+\log(z),~2\log(-z)=2\log(z)
\Rightarrow \log(-z)=\log(z)$; in particular $\log(-1)=\log(1)=0$ what is not true as $\log(-1) = i(2k+1)\pi\neq 0$).
Hence the above logarithmic integral for complex argument is defined as
${\Li}(x^\rho)={\Li}(e^{\rho\log(x)})$,  where for $z=u+iv, ~v\neq 0$:  
\bee
{\Li}(e^z)=\int_{-\infty+iv}^{u+iv\\~~~~} \frac{e^w}{w}dw,
\eee
thus  $\Li$ is in fact defined via  the exponential integral. Let us mention, that in Mathematica to obtain the value
of ${\Li}(x^{\rho_k})$ the command \verb"ExpIntegralEi[ZetaZero[k]*Log[x]]" has to be used.

In equations \eqref{J-explicite} and \eqref{pi-exact} we meet the issue of determining  zeros $\rho$ of the zeta function:
$\zeta(\rho)=0$.  Riemann has shown that $\zeta(s)$ fulfills the {\it  functional identity}:
\bee
\pi^{-\frac{s}{2}}\Gamma\left(\frac{s}{2}\right)\zeta(s)
           =\pi^{-\frac{1-s}{2}}\Gamma\left(\frac{1-s}{2}\right)\zeta(1-s),~~~~{\rm  for } ~ s \in\mathbb{C}\setminus\{0,1\}.
\label{functional-zeta}
\eee
The above form of the functional equation is explicitly symmetrical with respect to the line $\Re(s)=1/2$: the change
$s\rightarrow \frac{1}{2}+s$ on both sides of (\ref{functional-zeta}) shows that the values of the combination
of functions $\pi^{-\frac{s}{2}}\Gamma\left(\frac{s}{2}\right)\zeta(s)$  are  the same at points  $ \frac{1}{2}+s$  and
$\frac{1}{2}-s$.

\begin{figure*}[ht]
\centering
\includegraphics[width=0.75\textwidth, angle=0]{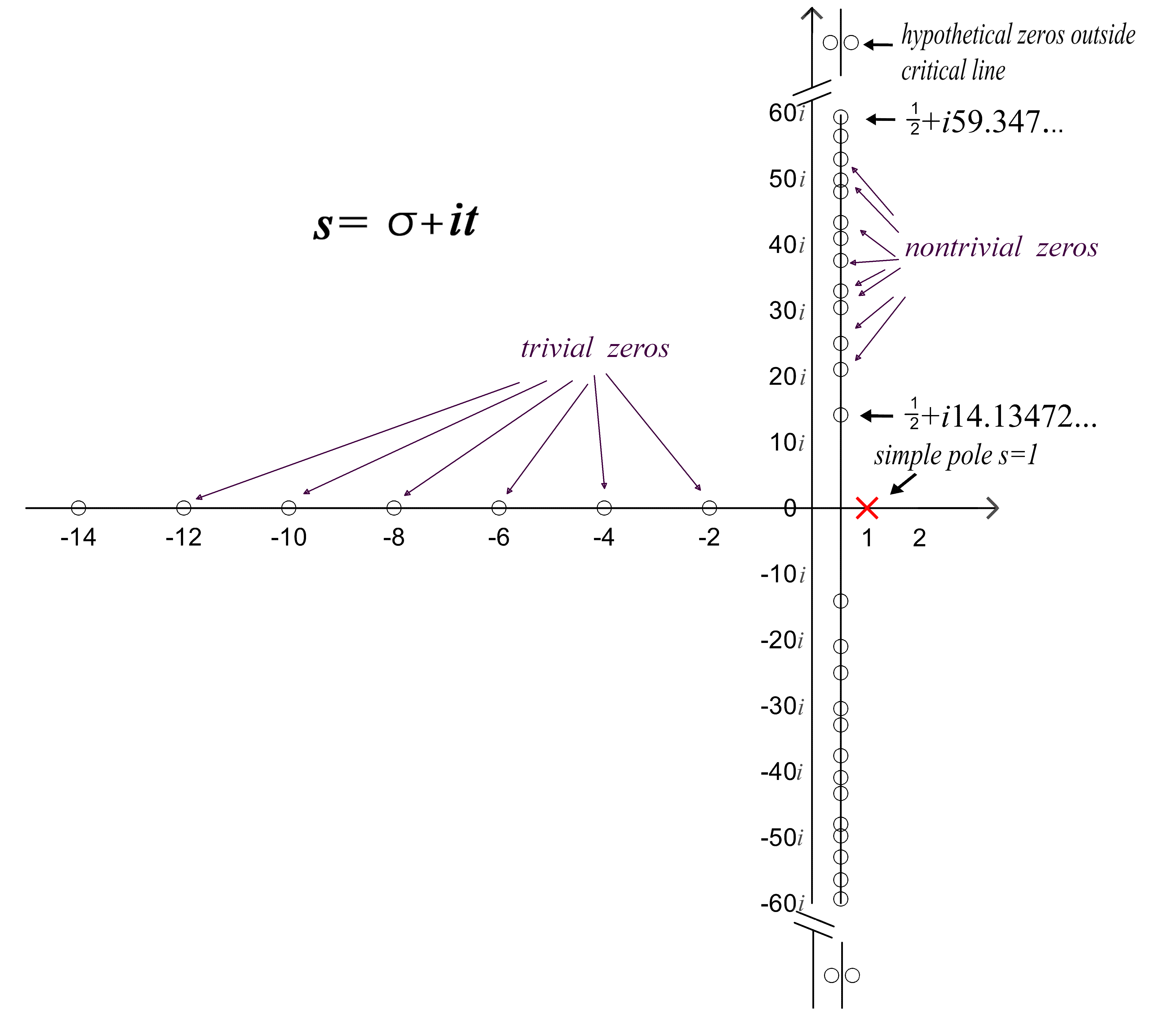}
\caption{\small The location of zeros of the Riemann $\zeta(s)$ function.}
\bigskip
\label{plane}
\end{figure*}

Because $\Gamma(z)$ is singular at all negative integers, thus to fulfill functional identity (\ref{functional-zeta})
$\zeta(s)$ has to be zero at all negative even integers:
\[
\zeta(-2n)=0,  n=1, 2, 3, \ldots
\]
These zeros are called {\it trivial} zeros.  The fact that $\zeta(s)\neq 0$ for $\Re(s)>1$ and the  shape of the  functional
identity entails that {\it nontrivial}  zeros $\rho_n=\beta_n + i\gamma_n$ are located in the {\it critical  strip}:
\[
0 \leq  \Re [\rho_n] =\beta_n \leq 1.
\]
From the symmetry of the functional equation (\ref{functional-zeta})  with respect to the line $\Re[s]=\frac{1}{2}$
it follows, that if  $\rho_n=\beta_n + i\gamma_n$ is a zero, then
$\overline{\rho_n}=\beta_n -i\gamma_n $ and  $1-\rho_n, ~1-\overline{\rho_n}$ are also zeros: they are located symmetrically
around  the straight line  $\Re[s]=\frac{1}{2}$ and the axis $t=0$, see Fig. \ref{plane}.

The sum over trivial zeros $\rho=-2n$ in \eqref{J-explicite} can be calculated analytically giving the {\it explicit}
(i.e. expressed directly by sum over zeros of $\zeta(s)$) formula for $J(x)$:
\bee
J(x)={\Li}(x)-\sum_{\substack{\Re(\rho)>0\\0<\Im(\rho)<1}}\left({\Li}(x^\rho)+{\Li}(x^{\overline{\rho}})\right)  +\int_x^\infty \frac{du}{u(u^2-1)\log(t)}-\log(2)
\eee
and therefore {\it explicit} formula  for $\pi(x)$ follows:
\bee
\pi(x)=\sum_{n=1}^N \frac{\mu(n)}{n}\left({\Li}(x^{\frac{1}{n}})-\sum_{\substack{\Re(\rho)>0\\0<\Im(\rho)<1}} \Bigl({\Li}(x^\frac{\rho}{n})+
{\Li}(x^\frac{\overline{\rho}}{n})\Bigr)  +\int_{x^{1/n}}^{\infty } \frac{1}{u \left(u^2-1\right)\log(u)} \, du - \log(2) \right).
\label{pi-exact-b}
\eee
Extending in the first term  summation to infinity (it is not an big sin, as terms with large $n$ tend to zero) gives the function
\bee
R(x)=\sum_{n=1}^\infty \frac{\mu(n)}{n} {\rm Li}(x^{1/n}).
\label{R-od-x}
\eee
The function $R(x)$ can be calculated from very quickly converging series
\bee
R(x)=1+\sum_{n=1}^\infty \frac{\left(\log(x)\right)^n}{n n!\zeta(n+1)}.
\label{R-od-x-Gram}
\eee
The last sum is called the Gram formula,  see  \cite[p.51]{Riesel}  for transformations leading from
(\ref{R-od-x}) to (\ref{R-od-x-Gram}).
Because the sum  over all complex zeros is not absolutely convergent  its value depends
on the order of summation.  In fact  famous (and curious) Riemann's rearrangement theorem,  see e.g.  \cite[Theorem 3.54] {Rudin-principles},
asserts that terms of a conditionally convergent infinite series  can be  permuted such a way  that the new series converges to       
{\it any given value}\ ! For \eqref{pi-exact-b} Riemann in \cite{Riemann-1859} says that  ``It may easily be shown, by means of
a more thorough discussion'' that the ''natural order'', i.e. the process of pairing together zeros $\rho$ and
$\overline{\rho}$ in order of increasing imaginary parts of $\rho$,  is the correct one.
At the end  of \cite{Riemann-1859} Riemann states  about the series in \eqref{pi-exact-b}  that ``when reordered it can take
on any arbitrary real value''.

\begin{figure*}[t]
\centering

\includegraphics[width=0.95\textwidth, angle=0]{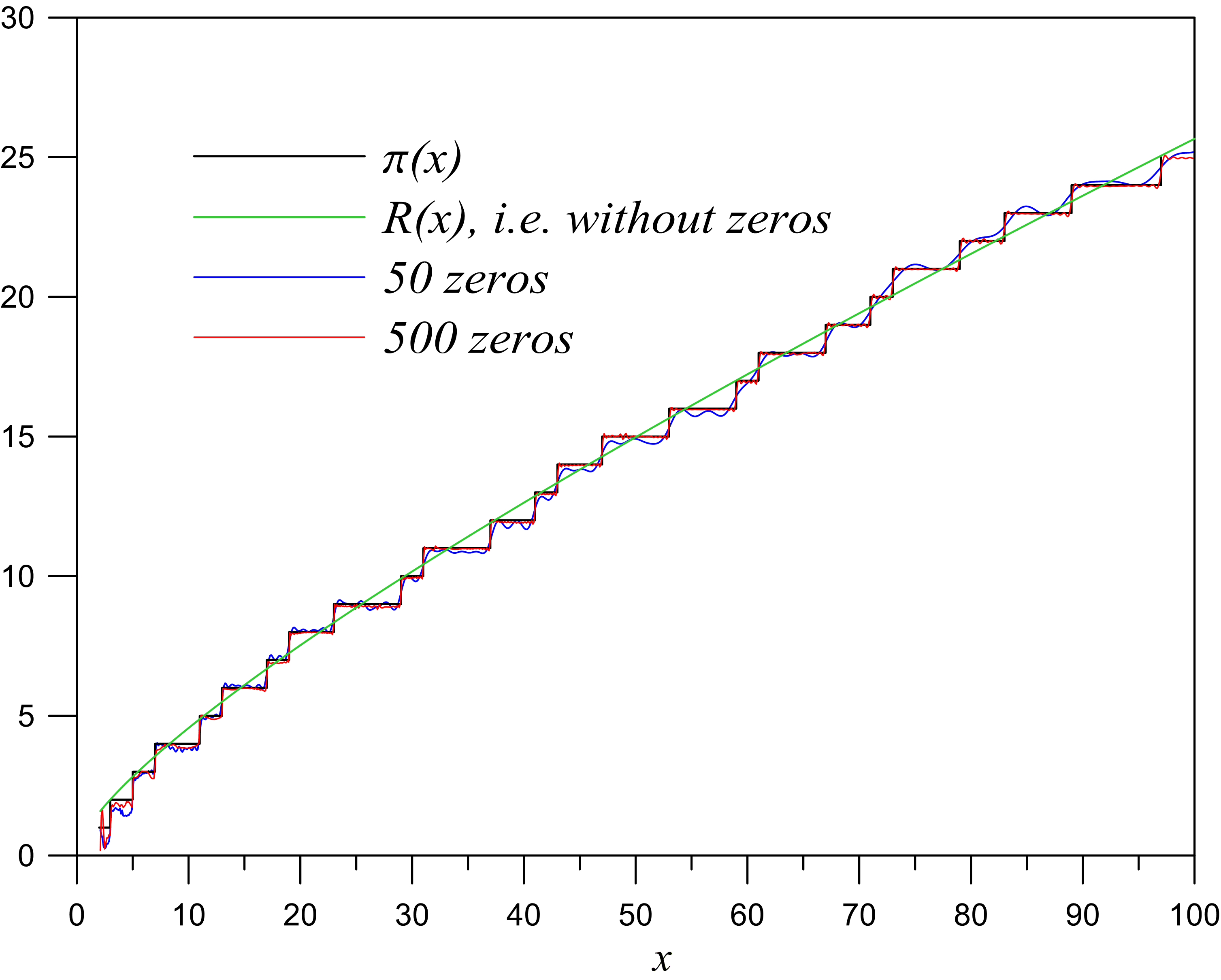} \\  
\caption{\small The plot of $\pi(x)$ and the r.h.s. of (\ref{pi-exact-b})  obtained by summing over first 50 nontrivial
zeros (blue line) and first 500 nontrivial zeros (red line). We have produced data for this plots using the equation \eqref{fale}.
The animation showing the effect of adding  consecutive zeros  one by one  to the formula
(\ref{pi-exact-b}) can be seen at http://empslocal.ex.ac.uk/people/staff/mrwatkin/zeta/encoding1.htm }
\label{pi-R}
\end{figure*}

Again let us point out the curiosity (mystery) of the above equation \eqref{pi-exact-b}: $\pi(x)$ on lhs jumps by 1 at each
argument being a prime with  constant values (horizontal sections) between two consecutive primes.  Thus on the rhs  the zeros
of zeta have to conspire to deform smooth plot of the first term ${\Li}(x)$ into the stair--like
graph with jumps.  It resembles the Fourier series of smooth sinuses reproducing say the step function on interval $(-\pi, \pi)$.
In Fig. \ref{pi-R} we have made plots  illustrating these observations.

The formula \eqref{pi-exact-b} is less time consuming to obtain $\pi(x)$ for large $x$ than counting all primes up to $x$;
the best non--analytical methods of computing $\pi(x)$ have complexity $\mathcal{O}(x^{2/3}/\log^2(x))$, while
involving  some variants of the Riemann explicite formula are  $\mathcal{O}(x^{1/2})$ in time.
For example, the value $\pi(10^{24})=18, 435, 599, 767, 349, 200, 867, 866 $  was obtained by a variant of \eqref{pi-exact-b}
using     $59, 778, 732, 700$ nontrivial zeros of $\zeta(s)$ \cite{Platt-pi}. Also the value
$\pi(10^{25})=176,846,309,399,143,769,411,680$ was announced, see \href{http://oeis.org/A006880}
{The On-Line Encyclopedia of Integer Sequence, entry A006880}.

Amazingly the horrible looking sum of the integrals in \eqref{pi-exact-b} stemming from the trivial zeros can be brought to
the  simple  closed form:
\bee
\sum_{n=1}^N \frac{\mu(n)}{n}\left( \int_{x^{1/n}}^{\infty } \frac{1}{u \left(u^2-1\right)\log(u)} \, du  - \log(2) \right)=
 \frac{1}{2\log x}\sum_{n=1}^N \mu(n) + \frac{1}{\pi} \arctan \frac{\pi}{\log x}+\epsilon(x, N),
 \label{Riesel}
\eee
where $\epsilon(x, N)\rightarrow 0$ as $N \rightarrow \infty$,  for details see \cite{Riese-Gohl}. The special choice of $N$ such that
$\sum_{k=1}^N \mu(k)=-2$ (e.g. $N=5, 7, 8, 9, 11, 12, \ldots$) is favoured: the series for arcus tangens in the vicinity of $u=0$
has the form $\arctan(u)=u-u^3/3+u^5/5-u^7/7+\ldots$  and for such a special $N$
the first two terms in \eqref{Riesel}  behave together like $(\pi/\log(x))^3/3+\ldots$ thus the contribution from trivial zeros is
negligible for large $x$ and hence nontrivial zeros are prevailing.

So where are the complex zeros of zeta?  Riemann has made the assumption, now called the {\it \bf Riemann Hypothesis},
that all nontrivial zeros lie on the {\it critical line} $\Re[s]=\frac{1}{2}$:
\bee
\rho_n=\frac{1}{2} + i\gamma_n  ~~~~~(\mbox{i.e. }~~ \beta_n=\frac{1}{2}~~ \mbox{for all} ~~n).
\eee
Contemporary the above requirement is augmented by the demand that all nontrivial zeros are simple.
Despite many efforts the Riemann Hypothesis remains unproved.
In  Fig.\ref{plane} we illustrate the Riemann Hypothesis and in the Table 1 we give the approximate values of
the first 10 non-trivial zeros  of $\zeta(s)$. We see that initially $\gamma_n>n$,  but at $n=9137$ the inequality
reverses as $\gamma_{9137}=9136.6792\ldots$ while  $\gamma_{9136}=9136.1396\ldots$. Next zero is again larger than its
index: $\gamma_{9138}=9138.10591\ldots$ but $\gamma_{9141}= 9140.5783\ldots$ and up to $n=10,000,000$ the inequality
$\gamma_n<n$ holds  and we believe it will be fulfilled for ever.  
Incidentally $9137$ is a prime,  in fact it is a left--truncatable prime: removing
successively left digit  gives again prime numbers 9137, 137, 37 and 7.  

\begin{center}
{\centerline{ \bf Table 1}}
\bigskip
\begin{tabular}{|c|c||c|c|} \hline
$ n  $  &  $ \tfrac{1}{2}+i\gamma_n $  &   $ n  $  &  $ \tfrac{1}{2}+i\gamma_n $    \\ \hline
$ ~~1~~~$ & ${\tfrac{1}{2}} + i       14.134725142\ldots ~$ & $ ~~6~~~$ & ${\tfrac{1}{2}} + i       37.586178159\ldots ~$  \\ \hline
$ ~~2~~~$ & ${\tfrac{1}{2}} + i       21.022039639\ldots ~$ & $ ~~7~~~$ & ${\tfrac{1}{2}} + i       40.918719012\ldots ~$  \\ \hline
$ ~~3~~~$ & ${\tfrac{1}{2}} + i       25.010857580\ldots ~$ & $ ~~8~~~$ & ${\tfrac{1}{2}} + i       43.327073281\ldots ~$  \\ \hline
$ ~~4~~~$ & ${\tfrac{1}{2}} + i       30.424876126\ldots ~$ & $ ~~9~~~$ & ${\tfrac{1}{2}} + i       48.005150881\ldots ~$  \\ \hline
$ ~~5~~~$ & ${\tfrac{1}{2}} + i       32.935061588\ldots ~$ & $ ~~10~~~$ & ${\tfrac{1}{2}} + i       49.773832478\ldots ~$ \\ \hline
\end{tabular}
\end{center}

\begin{figure*}[ht]
\centering
\includegraphics[width=0.9\textwidth, angle=0]{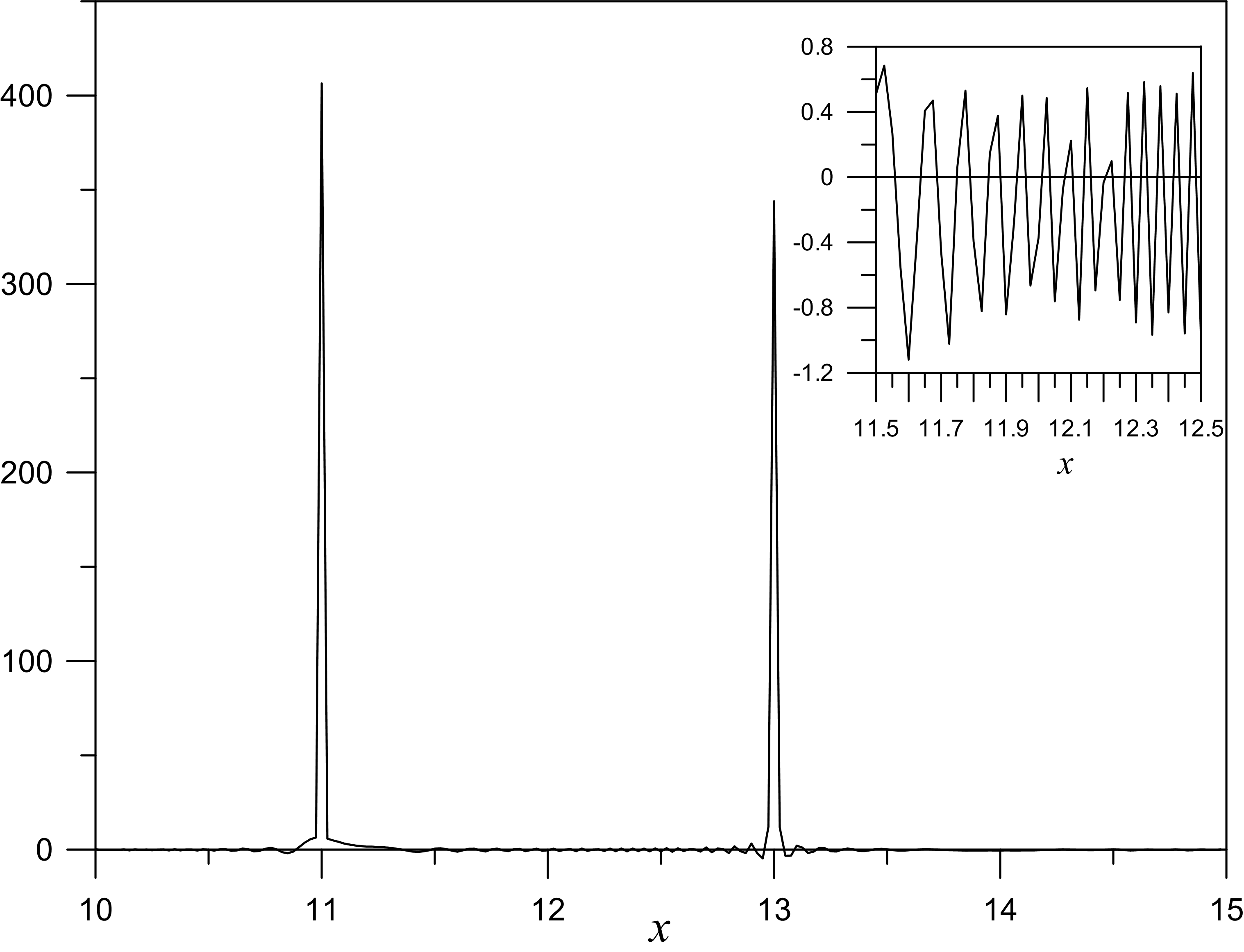} \\  
\caption{\small The plot of the derivative of two sums on rhs of  \eqref{fale} obtained by summing over first 15000 nontrivial
zeros showing delta--like pattern. In the inset fluctuations for $11.5<x<12.5$ are shown --- the high spikes
at $x=11$ and $13$  squeezed  them on the original plot.}
\label{Dirac}
\end{figure*}

Assuming the RH, i.e. collecting together terms $\rho_n=\frac{1}{2}+i\gamma_n$ and $\overline{\rho_n}=1-\rho_n=\frac{1}{2}-i\gamma_n$,
using the Euler identity $e^{i\alpha}=\cos(\alpha)+i\sin(\alpha)$ we can represent $\pi(x)$ as a main smooth trend plus
superposition of waves  $\sin(\cdot)$ and  $\cos(\cdot)$:
\bee
\begin{split}
\pi(x)=1+\sum_{n=1}^\infty \frac{(\log(x))^n}{n n! \zeta(n+1)}~-~~~~~~~~~~~~~~~~~~~~\\
\sum_n \mu(n)\frac{x^{\frac{1}{2n}}}{\log(x)}  \sum_i \Biggl\{  \frac{\cos\left(\frac{\gamma_i \log(x)}{n}\right)+2\gamma_i\sin\left(\frac{\gamma_i\log(x)}{n}\right)}{\frac{1}{4}+\gamma_i^2} \\
+ \frac{n}{\log(x)}\frac{(\frac{1}{4}-\gamma_i^2)2 \cos\left(\frac{\gamma_i \log(x)}{n}\right)+2\gamma_i\sin\left(\frac{\gamma_i\log(x)}{n}\right)}{\frac{1}{16}+\frac{1}{2}\gamma_i^2+\gamma_i^4}\Biggr\},  \\
\end{split}
\label{fale}
\eee
where we have used two first terms of the expansion ${\Li}(x)\approx x/\log(x)+x/\log^2(x)$.  Using the above equation  with  10000
zeros and second sum over $n$ up to 7  we obtained  $25.00267$ for $\pi(100)$, while the numbers of primes up to 100 (without counting 1 as
a prime)   is 25. Physicists  well know that derivative of the step function is the Dirac delta function: $\Theta'(x)=\delta(x)$,
thus the  derivative of $\pi(x)$  with respect to $x$ is  the  sum of  Dirac deltas concentrated on primes: $\sum_{p_n}\delta(x-p_n)$.
We have differentiated two first sums in \eqref{fale}, i.e. skipping terms $\mathcal{O}(1/\gamma_k^4)$,  summed over
first 15000 nontrivial zeros of zeta and the resulting plot is presented in Fig.\ref{Dirac}.
The animated  plot of the delta--like spikes emerging with increasing number of nontrivial zeros taken into account is available
at \href{http://empslocal.ex.ac.uk/people/staff/mrwatkin/zeta/pianim.htm}{http://empslocal.ex.ac.uk/people/staff/mrwatkin/zeta/pianim.htm}.

In 1896 J. S. Hadamard (1865 -- 1963) and  Ch. J. de la Vallée Poussin  (1866 -- 1962) independently proved that
$\zeta(s)$ does not have zeros on the line $1+it$,  thus $|x^\rho|<x$. It suffices to obtain from \eqref{pi-exact-b} the original
Gauss's  guess \eqref{PNT},  which thus became a theorem called the Prime Number Theorem (PNT). Indeed:  for large $x$
in \eqref{pi-exact-b} the first term $R(x)$ wins over terms with ${\Li}(x^\rho)$
and then from  \eqref{R-od-x} we have that $R(x)\approx {\Li}(x)$.

Already Riemann  calculated numerically a few first nontrivial zeros of $\zeta(s)$
\cite{Edwards}. Next in 1903 J.P. Gram \cite{Gram} calculated that first 15 zeros of $\zeta(s)$ are on the critical line;
in June of 1950 A.Turing has used the Mark 1 Electronic Computer at Manchester University to check that first  1104 zeros
are on the critical line. He has done this calculations ``in an optimistic hope that a zero would be  found off critical line'',
see \cite[p. 99]{Turing}. A few years ago S. Wedeniwski (2005)
was leading the  internet project Zetagrid \cite{Zetagrid} which during four years
determined that  $250\times 10^{12}$ zeros are on the critical line, i.e. on $s=\frac{1}{2}+it$ up to  $~t < 29,538,618,432.236$.
The present record belongs to  K. Gourdon(2004) \cite{Gourdon}: the first  $10^{13}$ zeros are on the critical line.
A.  Odlyzko checked that RH is true in different intervals
around $10^{20}$ \cite{Odlyzko_zera1}, $10^{21}$ \cite{Odlyzko_zera2}, $10^{22}$
\cite{Odlyzko2001}, but his aim was not to verify the RH but rather providing evidence
for conjectures that relate nontrivial zeros  of $\zeta(s)$ to eigenvalues of
random matrices, see Sect. \ref{rmt}.  In fact Odlyzko has expressed the view that the hypothetical zeros off
the critical line are unlikely to be encountered for $t$  below $10^{10^{10000}}$, see \cite[p.358]{Derbyshire}.

\begin{figure*}[h]
\hspace{0.03\textwidth}
\centering
\includegraphics[width=0.9\textwidth]{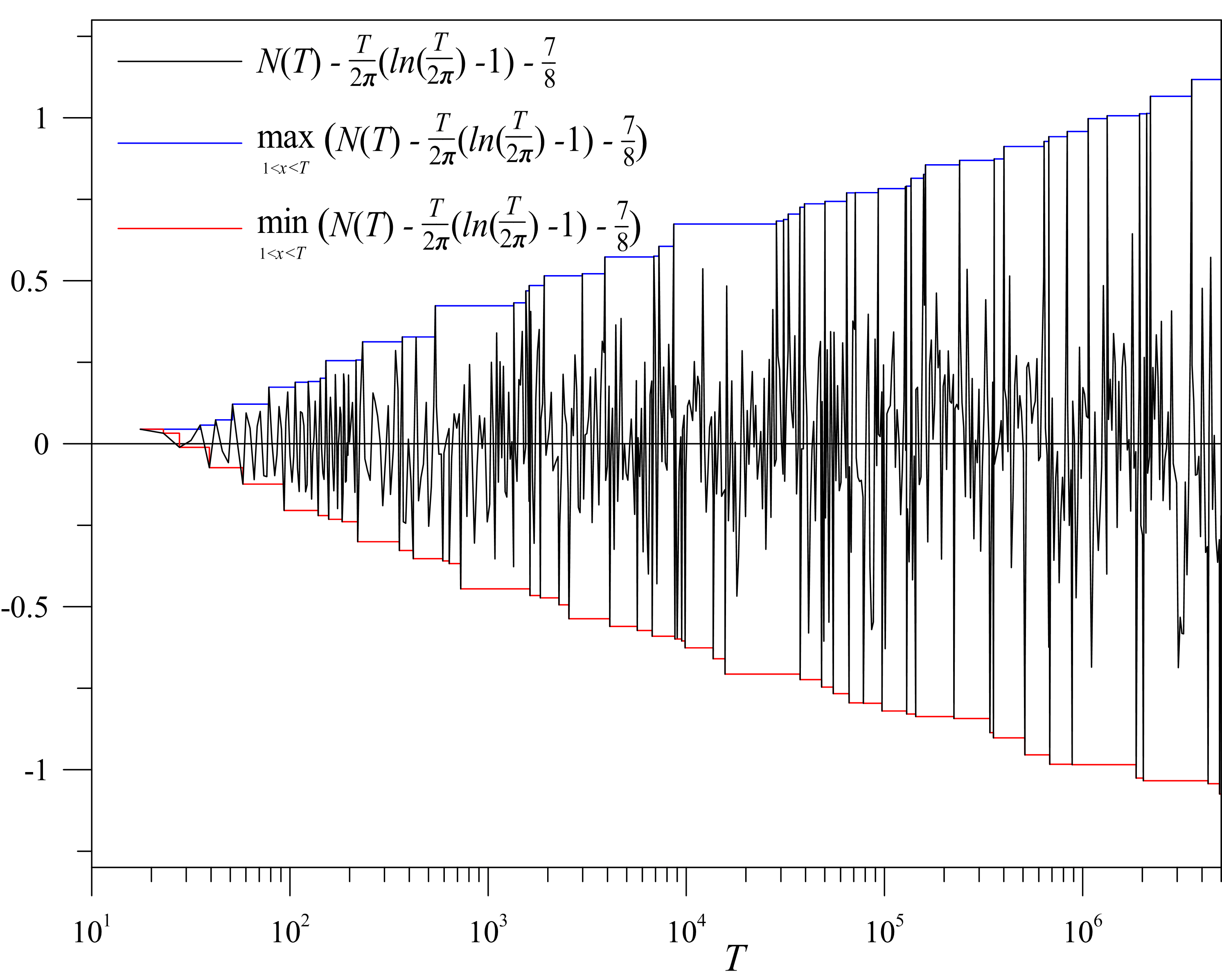}
\caption{\small The plot illustrating the formula  \eqref{von-Mangoldt} for number of nontrivial zeros up to $T=5\times 10^6$. When on vertical
axis the interval (0,1) is 4 cm long, then if the horizontal  axis would be plotted
on linear  scale instead of logarithmic, its length should amount to 200  km. This analogy in a striking way shows how precise
is the formula \eqref{von-Mangoldt} and that the term $7/8$ is important (mathematicians will not agree with this and skip this
constant term).  }
\label{von-Mangoldt-figure}
\vspace{0.3cm}
\end{figure*}

Let  $N(t)$ denote the function counting the nontrivial zeros up to $T$, i.e. $N(T)=\sum_n \Theta(T-\Im[\rho_n])$.
In his seminal paper  Riemann announced and in 1905 von Mangoldt proved that:
\bee
N(T)=\frac{T}{2\pi}\log\left(\frac{T}{2\pi e}\right)+\frac{7}{8} +{\mathcal{O}}(\log(T)).
\label{von-Mangoldt}
\eee
The Fig.\ref{von-Mangoldt-figure} illustrates how well the above formula predicts $N(T)$.
In 1904 G.H. Hardy proved, by considering  moments of certain functions related to the zeta function,  that on the critical line
there is infinity of zeros of  $\zeta(s)$ \cite{Hardy1914}.  Levinson (1974) proved that more than one-third of
zeros  of  Riemann's $\zeta(s)$ are  on critical line by relating
the zeros of the zeta function to those of its derivative, and Conrey (1989) improved this further to two-fifths
(precisley 40.77 \%   have $\Re(\rho)=\tfrac{1}{2}$).
The present record  seems to belong to S. Feng, who proved that at least 41.28\% of the zeros of the Riemann zeta
function are on the   critical line \cite{Feng}.

At the end of this Section we mention, that $\zeta(s)$ admits besides the product \eqref{gold-formula} another product
representation, called the Hadamard product:
\bee
\zeta(s)=\frac{\pi}{(s-1) \Gamma\left(\frac{s}{2}-1\right)}  e^{-1-\frac{\gamma}{2}}  \prod_{k=1}^{\infty } e^{\frac{s}{\rho_k}} \left(1-\frac{s}{\rho _k}\right).   
\eee
In contrast to \eqref{gold-formula} it is valid on the whole complex plane without $s=1$.
It is an example of the general  Weierstrass factorization theorem:  points where function  vanishes determine this function.
We also  add that   the common belief is  that the imaginary parts $\gamma_l$ of the nontrivial zeros of $\zeta(s)$
are irrational and perhaps even transcendental \cite{Odlyzko-chaos}, \cite{Wolf-zeta-zeros}.

\section{How to prove the Riemann Hypothesis?}
\label{jak}

Practically nobody is going to prove the RH directly: there are probably well over one hundred of different facts either
equivalent to RH or of whose truth RH will follow (i.e. sufficient conditions). Hence proving one of these so  called
criteria for RH will entail the validity of RH. Below we present a few such criteria for RH.

In 1901 H. von  Koch  proved \cite{Koch} that the Riemann Hypothesis is equivalent to the following error term
for the  approximation of the prime counting function by logarithmic integral:
\bee
\pi(x) = {\rm Li} (x) + {\mathcal{O}}(\sqrt{x}\log(x)).
\eee
Later the error term was specified explicitly by Schoenfeld \cite[Corollary 1]{Schoenfeld1976}  and RH is equivalent to
\bee
|\pi(x) - {\rm Li} (x)| \leq \frac{1}{8\pi}\sqrt{x}\log(x)\mbox{  for all  } x\geq 2657.
\label{Koch}
\eee

The following facts show that the validity of the RH is very delicate and subtle: namely in some sense RH is valid with
accuracy $\epsilon=1.14541 \times 10^{-11}$  (or less, that is the present  value of  $\epsilon$).  Here is the reasoning
leading to this  conclusion: Let us introduce the function
\bee
\xi(iz)=\frac{1}{2}\left(z^2-\frac{1}{4}\right)\pi^{-\frac{z}{2}-\frac{1}{4}}\Gamma\left(\frac{z}{2}+\frac{1}{4}\right)
\zeta\left(z+\frac{1}{2}\right).
\eee
We can see from  the above formula that:  RH is true  $\Leftrightarrow$ all zeros of $\xi(iz)$
are real. The point is that $\xi(z)$ can be expressed as the following Fourier transform (for derivation of this formula
see  e.g. \cite[Sect.10.1]{Titchmarsh}):
\bee
\frac{1}{8}\xi\left(\frac{z}{2}\right)=\int_0^\infty \Phi(t)\cos(zt) dt,
\eee
where
\bee
\Phi(t)=\sum_{n=1}^\infty (2\pi^2 n^4 e^{9t} - 3\pi n^2 e^{5t}) e^{-\pi n^2e^{4t}}.
\eee
The functin $\xi(z)$ can be generalized  to the family of functions  $H(z,\lambda)$ parameterized by $\lambda$:
\bee
H(z, \lambda) =\int_0^\infty \Phi(t)e^{\lambda t^2} \cos(zt) dt.
\eee
Thus we have $H(z,0)=\frac{1}{8} \xi(\frac{1}{2}z)$.  N. G. De Bruijn proved  in 1950 \cite{de_Bruijn} that $H(z,\lambda)$ has
only real zeros for $\lambda\geq \frac{1}{2}$  and  if $H(z,\lambda)$ has only real zeros for some  $\lambda'$, then
$H(z,\lambda)$ has only real zeros for each $\lambda>\lambda'$. In 1976 Ch. Newman  \cite{Newman} has proved that there
exists parameter $\lambda_1$ such that  $H(z,\lambda_1)$ has at least one non-real zero. Thus, there exists
such constant $\Lambda$ in the interval $-\infty < \Lambda <\frac{1}{2}$  that  $H(z,\lambda)$ has real zeros
$\Leftrightarrow \lambda>\Lambda$.  The Riemann  Hypothesis is equivalent to $\Lambda\leq 0$. This constant $\Lambda$ is now
called the de Bruijn--Newman constant.  Newman believes that  $\Lambda \geq 0 $. The computer determination has
provided the numerical estimations of values of de Bruijn--Newman constant; the current record belongs to 
Y. Saouter {\it et. al.} \cite{Saouter-et-al}:   $ \Lambda > -1.14541 \times 10^{-11}$.  Because the gap  in which $\Lambda$ catching the RH is so squeezed, Odlyzko noted
in \cite{Odlyzko2000}, that ``...the Riemann Hypothesis, if true,  is just barely true''.

There are also criteria for RH involving integrals. V. V. Volchkov has proved \cite{Volchkov95}  that following equality
is equivalent to RH:
\bee
\int_0^\infty \frac{1-12t^2}{(1+4t^2)^3}\int_{\frac{1}{2}}^\infty \log(|\zeta(\sigma+it)|d\sigma dt=\pi \frac{3-\gamma}{32}.
\eee
In the paper \cite{He2015jla} the above  integral  was used to express the RH in terms of the Veneziano amplitude for strings as well
as to find some generalizations of the Volchkov's  criterion.

In the paper \cite{Balazard-et-al} the equality to zero of the following integral was shown to be equivalent to RH:
\bee
\int_{\Re(s)=\frac{1}{2}}\frac{\log(|\zeta(s)|)}{|s|^2}|ds|=\int_{-\infty}^\infty \frac{\log(|\zeta(\tfrac{1}{2}+it)|)}{\tfrac{1}{4}+t^2} dt = 0.
\eee

\begin{figure*}[h]
\vspace{-0.3cm}
\begin{center}
\includegraphics[width=0.9\textwidth]{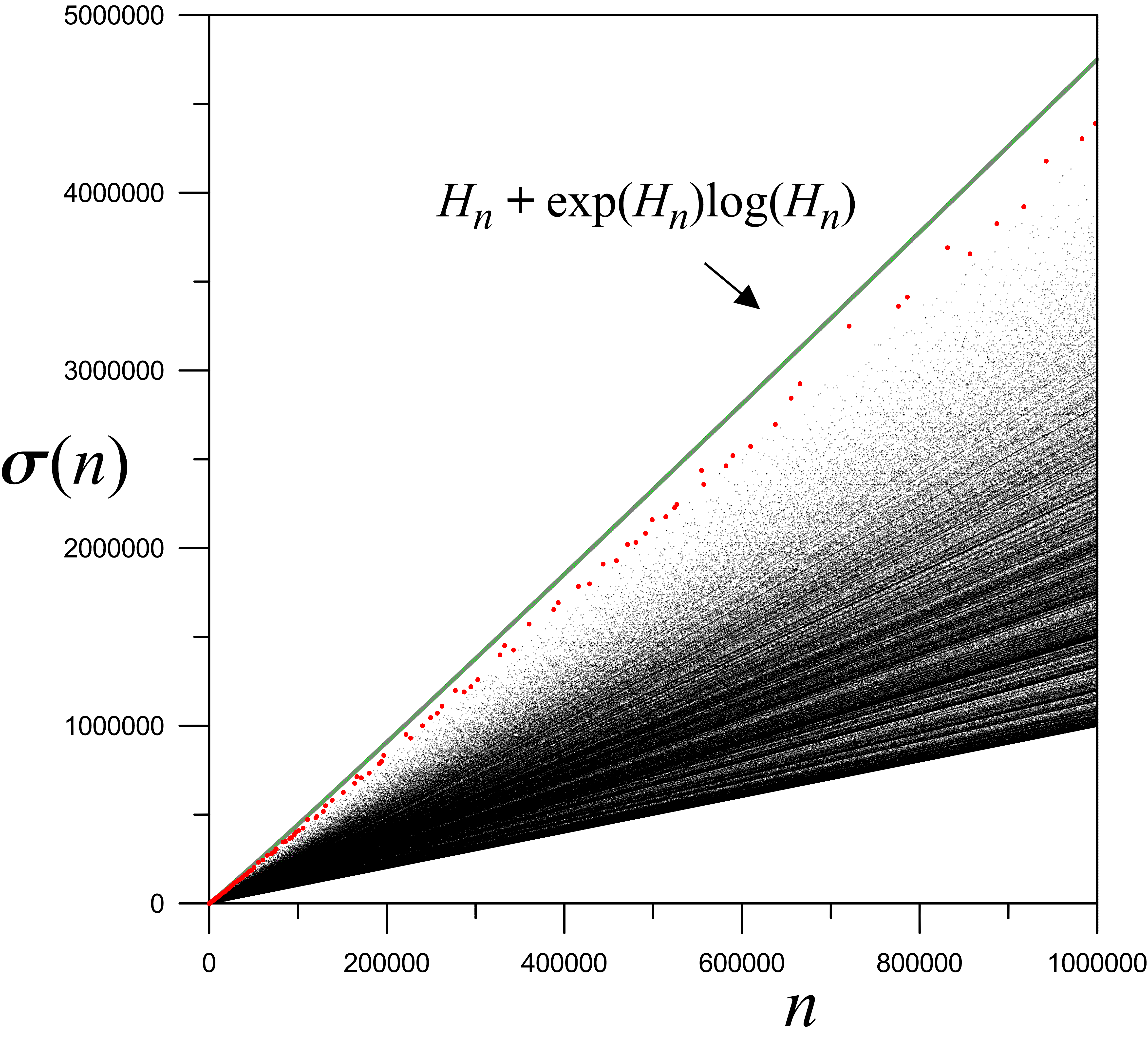} \\
\caption{\small  The plot of $\sigma(n)$ for $1<n<10^6$. In red are plotted values of
$\sigma(n)$ which  approach the threshold (green  line)  values closer than 10\%. The lower ``support''  of the graph  comes
from $n$ being   primes, as  prime $p$  is divisible only  by $1$ and itself,  so  $\sigma(p)=p+1$.  Data for this plot was
obtained  with the free package PARI/GP \cite{Pari}.}
\end{center}
\label{fig-lagarias}
\vspace{0.3cm}
\end{figure*}

Finally let us mention the elementary Lagarias criterion \cite{Lagarias}: the Riemann Hypothesis is equivalent to the  inequalities:
\bee
\sigma(n) \equiv \sum_{d|n} d \leq H_n + {\rm{e}}^{H_n}\log(H_n)
\label{Lagarias}
\eee
for each  $n=1,2,\ldots$,  where  $\sigma(n)$ is the sum of all divisors of $n$ and  $H_n$ is the $n$-th harmonic number
$ H_n=\sum_{j=1}^n \frac{1}{j}$. To disprove the RH it suffices to find one $n$ violating the inequality \eqref{Lagarias}.
The  Lagarias criterion is not well suited for computer verification (it is not an easy
task to calculate  $H_n$ for $n \sim 10^{100000}$ with sufficient accuracy)
and in \cite{Briggs2006}  K. Briggs has undertaken instead the verification of the Robin \cite{Robin}
criterion for RH:
\bee
{\rm RH~~} \Leftrightarrow  \sum_{d|n} d < e^\gamma n \log\log(n)~~~~{\rm for~~ }n>5040.
\eee
For some  $n$ Briggs obtained for the difference between r.h.s. and l.h.s. of the above inequality
value as small as  $e^{-13}\approx 2.2\times 10^{-6}$, hence again RH is in a danger to be violated.
As A. Ivic has put it  ``The Riemann Hypothesis is a very delicate mechanism.',  quoted in \cite[p. 123]{Sabbagh-zeros}.

Let us notice that the belief in  the validity of RH is not common:   famous mathematicians J. E. Littlewood,
P. Turan and A.M. Turing have  believed that the  RH is not true, see the paper ``On some reasons for doubting the Riemann
hypothesis'' \cite{Ivic2003}  (reprinted in \cite[p.137]{Borwein_RH})  written by  A. Ivi{\'c}, one of the
present day leading expert on RH. When  J. Derbyshire asked A. Odlyzko about his opinion on the validity of RH
he replied ``Either it's true, or else it isn't'' \cite[p. 357--358]{Derbyshire}.

\section{Quantum  Mechanics and RH}
\label{rmt}

The first physical method of proving RH was proposed by George Polya around 1914 during the conversation with Edmund Landau
and now is known as the Hilbert-Polya Conjecture.   Landau asked Polya:
``Do you know a physical reason that the Riemann hypothesis should be true?'' and  his reply was:
``This would be the case, I answered, if the nontrivial zeros of the $\Xi$-function were so connected with the physical
problem that the Riemann hypothesis would be equivalent to the fact that all the eigenvalues of the physical problem are real'',
\footnote{appearing here the function $\Xi$ is equal to the lhs of \eqref{functional-zeta} multiplied  by $s(s-1)/2$,
hence it has the same zeros as $\zeta(s)$}
see the whole story at the web site \cite{Correspondence}. Let us stress that this talk took place  many years before
the birth of quantum mechanics and the Schroedinger equation for energy levels. However in the period  1911--1914 Hermann   Weyl
published a few papers on the asymptotic distribution of eigenvalues of the Laplacian in the compact domain (in particular
the  eigenfrequencies or natural vibrations of the drums), see e.g.
\cite{Weyl1, Weyl2}. Thus presumably Polya was inspired by the Weyl's   papers.
If the RH is true nontrivial zeros lie on critical line and it makes sense to order them according to the imaginary part and eventually
put them  into the 1-1 correspondence with the eigenvalues of some hermitian operator. Therefore the problem is to find a connection
between  energy levels $E_n$ of some quantum  system and zeros of $\zeta(s)$.

\begin{figure*}[ht]
\begin{minipage}[ht]{0.5\textwidth}
\centering
\vspace{0.5cm}
\includegraphics[width=\textwidth, angle=0]{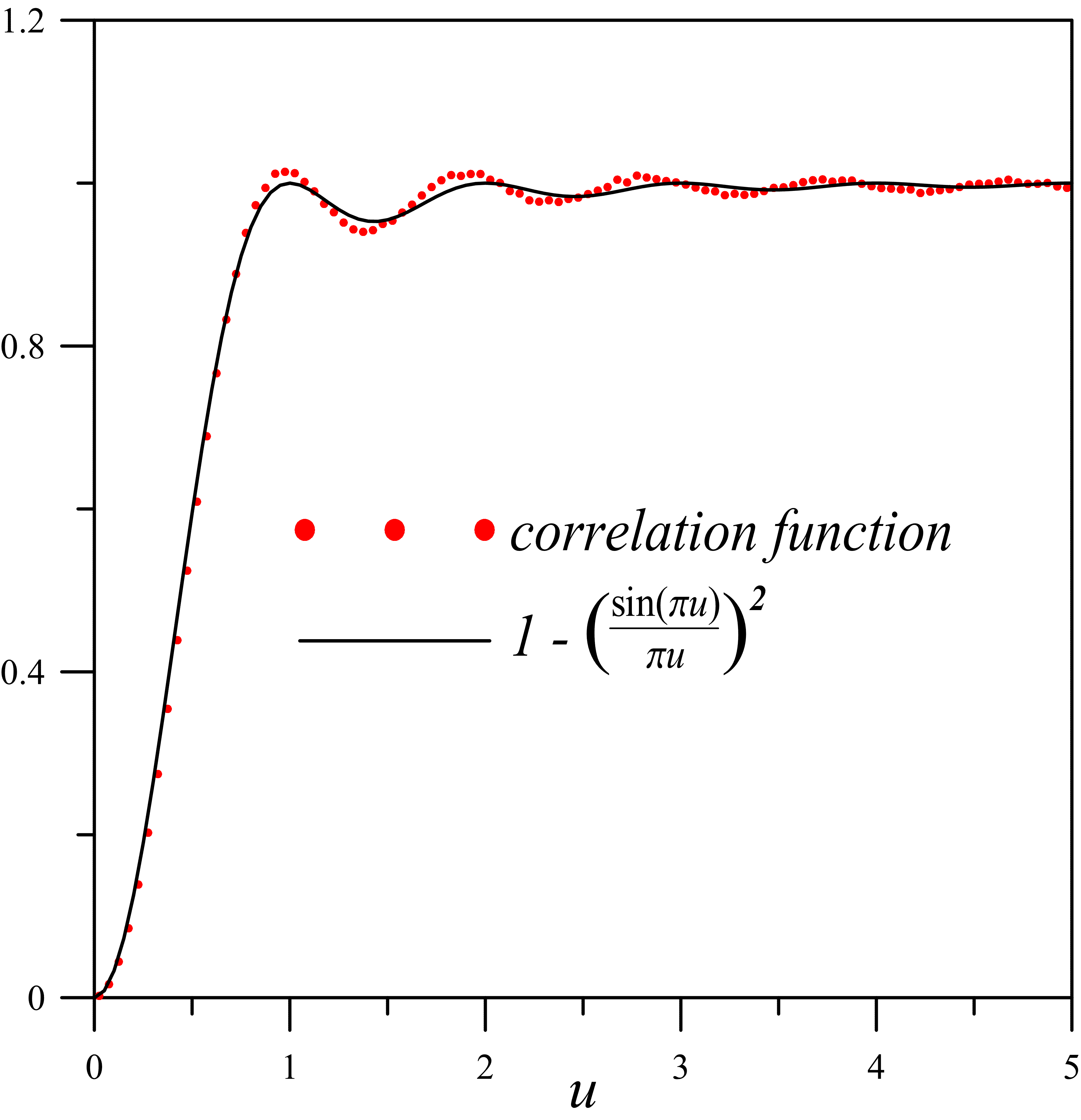} \\
\caption{\small  The plot of lhs of \eqref{correlations} calculated from the first $10^7$ zeros compared with the prediction of
Montgomery.  Points representing correlation function were calculated from equation on lhs of  \eqref{correlations}
for 5000000  first  zeros of  $\zeta(s)$  and  $\Delta  u = 0.05$.}
\label{fig-Montgomery}
\end{minipage}
\hspace{0.03\textwidth}
\begin{minipage}[ht]{0.5\textwidth}
\includegraphics[width=\textwidth,angle=0]{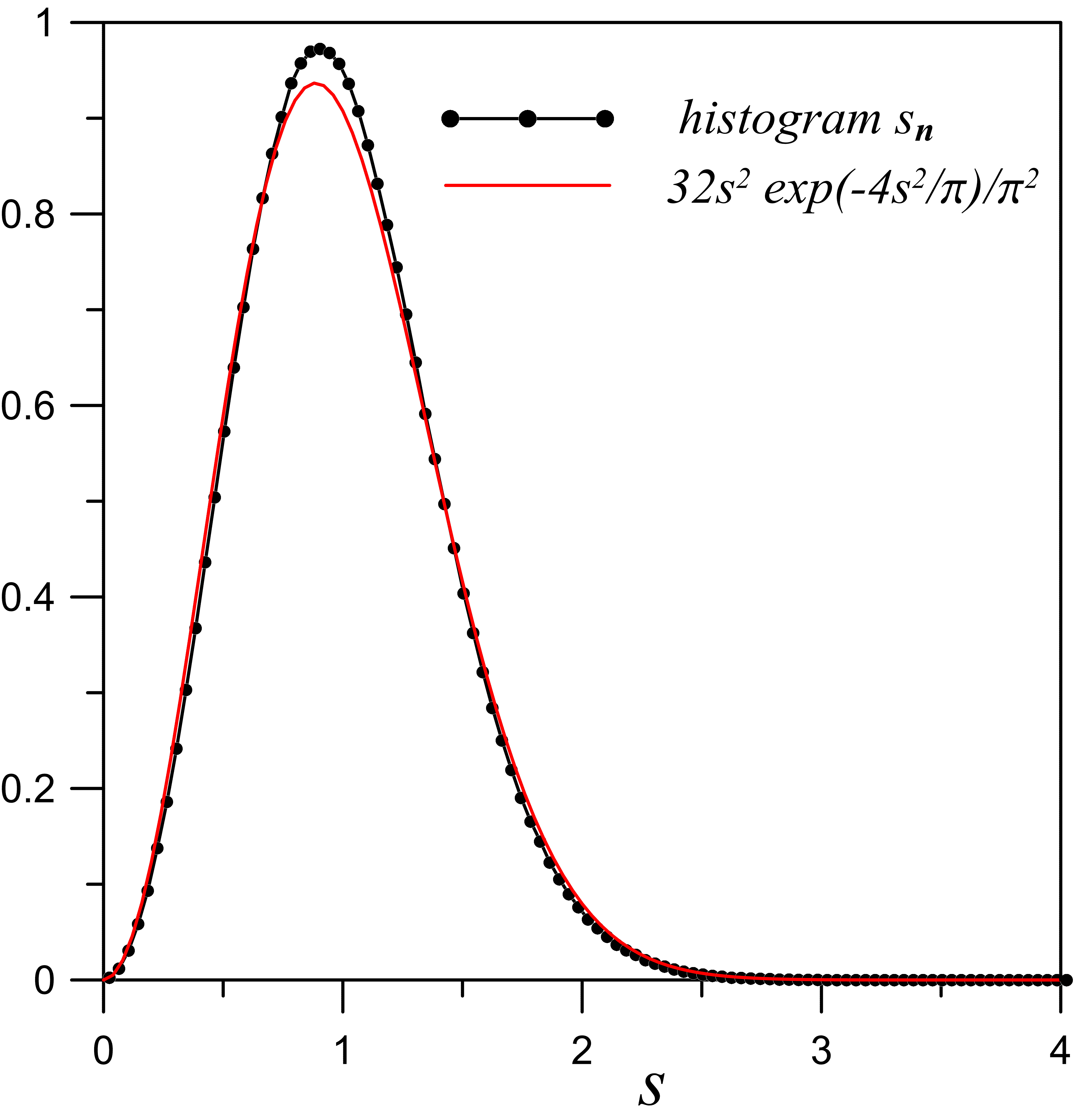}
\caption{\small Histogram  of normalized gaps between   cinsecutive zeros of  zeta obtained for   5000000  zeros.
Points are spaced with  $\Delta s =0.05$.  }
\label{fig-spacings}
\end{minipage}
\end{figure*}

In the autumn of 1971  \cite[p.261]{Sautoy2003}  H. Montgomery, assuming the RH, has proved   theorem about  statistical properties of
the spacings between zeta zeros. The formulation of this  theorem is rather complicated  and  we will not present it here,  see
his paper \cite{Montgomery-1973}. Next Montgomery made the conjecture  that  correlation function of the
imaginary  parts of nontrivial zeros has the form (here $0<\alpha<\beta<\infty$ are fixed):
\bee
\sum_{\substack{0<\gamma, \gamma '\leq T\\ \frac{2\pi \alpha}{\log T}   \leq\gamma - \gamma'
\leq  \frac{2\pi \beta}{\log T}}} 1  \rightarrow  \int_\alpha^\beta \left(1-\left(\frac{\sin\pi u}{\pi u}\right)^2\right )du~~~~{\rm as~~} T\rightarrow \infty .
\label{correlations}
\eee
In the Fig.\ref{fig-Montgomery}  we present a sketchy plot of the both sides of the equation \eqref{correlations}.
This result says  that the zeros –-- unlike primes, where it is conjectured that there is infinity of twins primes, i.e. primes
separated by 2, like $(3,5), (5,7), (11,13), \ldots, (59, 61), \ldots$)--- would actually repel one another because
in the integrand  $\sin(\pi u)/\pi u \rightarrow 1$  for  $u \rightarrow  0$.
Montgomery  published this result in 1973 \cite{Montgomery-1973},  but earlier in 1972 he spoke about
it with F. Dyson in Princeton,  see  many accounts of this story in the popular books listed in the Introduction,
e.g.  \cite[p.133--134]{Sautoy2003}. Dyson recognized  in \eqref{correlations}
the same dependence as in the behavior of the differences between pairs of eigenvalues of random Hermitian matrices.
The random matrices were introduced into the physics by Eugene   Wigner in the fifties of twenty century 
to model the energy levels in the nuclei  of  heavy atoms.  Spectra of light atoms are regular and simple in contrast
to the spectra of heavy atomic nuclei,  like e.g. $^{238}$U, for which hundreds of spectral lines were measured.
The hamiltonians of these nuclei  are not known,
besides that such many body systems are too complicated for analytical treatment. Hence the idea to model heavy nuclei
by  the matrix with random entries chosen according to the gaussian ensemble and subjected to some symmetry condition
(hermiticity etc.).

Because the hamiltonian describing  interaction  inside  heavy  nuclei is unknown Wigner proposed to  use  some matrix
of  large dimension with random    entries  selected with the appropriate distribution probability and subject
for example to the hermiticity requirement.  It means that if   ${\bf M}$ is a square matrix    $N\times  N$ with elements $M_{ij}$,
then probability  $P(M_{ij}\in (a,b))$  that a given matrix element  $M_{ij}$ will take value in the interval $(a, b)$
is given by the integral:
\[
P(M_{ij}\in (a,b))=\int_a^b f_{ij}(x) dx,
\]
where  $f_{ij}$ is the density of the probability distribution and matrix elements  $M_{ij}$ are  mutually statistically
independent, what means that the probability for the whole matrix is the product of above factors for single elements  $M_{ij}$.
The requirement of hermiticity (${\bf M}^\dagger ={\bf M}$) and independence with respect to the choice of the base determine
the following form, see \cite[Theorem 2.6.3,  p. 47]{Mehta}:
\bee
P({\bf M})=e^{-a \text{tr} \,{\bf M}^2 +b\text{tr}\,{\bf M} +c},
\label{postac-P}
\eee
where  $a$ is a positive real number,  $b$ i $c$ are real and  tr denotes trace of the matrix: $\text{tr}~{\bf M}=\sum_{i=1}^N M_{ii}$.
The value of  $c$  is determined by  normalization of the probability.   For self--adjoint matrix  ${\bf H}^\dagger = {\bf H }$
we have  $\text{tr} \,H^2  =  \sum_{i=1}^N \sum_{k=1}^N H_{ik}H_{ki}= \sum_{i=1}^N \sum_{k=1}^N H_{ik}H_{ik}^\star=
\sum_{i=1}^N \sum_{k=1}^N |H_{ik}|^2$  and all terms in   \eqref{postac-P} have a Gaussian form.  Because
exponent of the sum of terms is the  product of the  exponents of each factors separately,
the right side of the equation \eqref{postac-P} indeed has the form of the product of the density of the normal Gaussian
distributions and such a set of random,   Gaussian unitary  matrices  is  called  Gaussian Unitary Ensemble,
in short GUE.  Eigenvalues of such matrices are  not completely random:  ``unfolded''  gaps $s$ between  them
are not described by the Poisson  distribution  $e^{-s}$,  but for example for GUE  by the formula
\bee
    P(s) = \frac {32}{\pi^2} s^2 e^{-4 s^2/\pi}.
    \label{odstepy}
\eee
Unfolding means   getting rid of constant trend in the spectrum  $E_1, E_2, \ldots$,  i.e.  dividing $d_n=E_{n+1}-E_n$
by mean gaps between  levels $\overline{d}(E)$ :  $s_n =(E_{n+1}-E_n))/\overline{d}(E)$.  For zeros of
$\zeta(s)$  from equation   \eqref{von-Mangoldt} the  differences   $\gamma_{n+1}-\gamma_n$  are changed into
$s_n=(\gamma_{n+1}-\gamma_n)\log(\gamma_n)/2\pi$.  In the Fig.\ref{fig-spacings} we show the comparison of  \eqref{odstepy}
with real gaps for zeros of  $\zeta(s)$.

Level-spacing distributions of quantum systems can be grouped into a few universality classes connected
with the symmetry properties of the Hamiltonians: Poisson distribution for systems with underlying regular
classical dynamics, Gaussian orthogonal ensemble (GOE, also called the Wigner–-Dyson distribution) --- Hamiltonians
invariant under time reversal, Gaussian unitary ensemble  (GUE) --- not invariant under time reversal and Gaussian
symplectic ensemble (GSE) for half-spin systems with time reversal symmetry. There are many reviews on
these topics, we cite here \cite{Mehta}, \cite{Haake}, \cite{Weidenmuller-Mitchell}, we strongly recommend the review  
\cite{Firk-Miller}.   Dyson and Mehta identified these  three types of random matrices with  different intensities of repulsion
spacings between consecutive energy levels : GOE with weakest repulsion
between neighboring levels, GUE  with medium repulsion and GSE with strongest repulsion.   For quantitative description see
\cite[Appendix A]{Bohigas1989}.

For several years discovered  during a brief conversation of Montgomery with Dyson relationship  of  nontrivial zeros
$\zeta(s)$  with  the eigenvalues of matrix from the GUE did not arouse much interest.
In the eighties of previous century  Andrew Odlyzko  performed  over many years computation of zeta zeros  in different intervals 
and calculated their pair-correlation function numerically.  In the first paper \cite{Odlyzko1987} he tested Montgomery  pair  correlation
conjecture  for first 100,000 zeros and for zeros number $10^{12}$ to $10^{12}+100,000$. Next he looked at $10^{20}$-th zero
of the Riemann zeta function and 70 million of its neighbors, the $10^{20}$-th zero of the Riemann zeta function and 175
million of its neighbors, last  searched interval was around zero $10^{22}$ and involved $10^9$ zeros, see \cite{Odlyzko2001}.
The reason Odlyzko investigated zeros further and further is the very slow convergence of various characteristics of $\zeta(s)$
to its asymptotic behavior.
The results confirmed the GUE distribution:  the gaps between imaginary parts of consecutive nontrivial zeros of $\zeta(s)$
display  the same  behavior as  the differences between pairs of eigenvalues of random Hermitian matrices, see
\cite[Fig.1 and Fig.2]{Odlyzko1987}. In \cite[p.146]{Sarnak-1997b}
Peter Sarnak wrote: ``At the phenomenological level this is perhaps the most striking discovery about the zeta function since
Riemann.''  In this way vague hypothesis of Hilbert - Polya has gained credibility and now it is known that a physical system
corresponding to $\zeta(s) $ has to break  the symmetry with respect to time reversal.
At the conference ``Quantum chaos and statistical nuclear physics'' held in Cuernavaca, Mexico, in January 1986 Michael Berry delivered the
lecture {\it Riemann's zeta function: a model for quantum chaos?}  \cite{Berry-1986} which became the manifesto of the  approach
to prove the RH which can be summarized symbolically as  $\zeta(\frac{1}{2}+i\hat{H}_R)=0$ with $\hat{H}_R$ a hermitian operator
having as eigenvalues imaginary parts of nontrivial zeros $\gamma_k$:  $\hat{H}_R |\Psi_k\rangle=\gamma_k |\Psi_k\rangle$.
The hypothetical quantum system  (fictitious  element) described by
such a hamiltonian was dubbed by Oriol  Bohigas  ``Riemannium'',  see \cite{Leboeuf-et-al-2001, Hayes-2003}.
Additionally to the lack of time reversal invariance of  $\hat{H}_R$ Berry in \cite{Berry-1986} pointed out that  $\hat{H}_R$
should have a classical limit with classical orbits which are all chaotic (unstable and bounded trajectories). In fact the departure
of correlation function for zeta zeros from  \eqref{correlations} for large spacings (argument larger than 1 in
\cite[Fig.1 and Fig.2]{Odlyzko1987})  was a manifestation of quantum chaos,  as Berry recognized.
Later on M. Berry and J. Keating have argued \cite{Berry-Keating-1999} that $\hat{H}_R=xp$.  The main
argument for connection of   $\hat{H}_R=xp$ with the RH was the fact, that the number of states of this hamiltonian with
energy less than $E$ is given by the formula:
\[
N(E) = \frac{E}{2\pi}\left(\log\left(\frac{E}{2\pi}\right)-1\right)+\frac{7}{8}+\ldots
\]
what exactly coincides with \eqref{von-Mangoldt}. In the derivation of above result  Berry and Keating ``cheated''  using
very special Planck cell regularization to avoid infinite phase--space  volume.  As a caution we mention here an example of
a very special shape billiard  for which the formula for a number of energy levels below $E$ has a leading term
exactly the same as for zeta function \eqref{von-Mangoldt} but the next terms disagree,  see \cite[eqs. (34--35)]{Steiner-1990}.
We remind here that  two drums can  have different shapes but identical eigenvalues of vibrations, thus the same spectral staircase
function. In 2011 S. Endres and F. Steiner  \cite{Endres-Steiner}  showed that spectrum of  $\hat{H}_R=xp$ on the positive $x$ axis
is purely continuous and thus $\hat{H}_R=xp$ cannot yield the hypothetical Hilbert–-Polya  operator possessing as eigenvalues
the nontrivial zeros of the $\zeta(s)$ function.  The choice $\hat{H}_R=xp$  for  the  operator of  {\it Riemannium}
possesses  some additional drawbacks (e.g. it is integrable, and therefore not chaotic) and some modification of it were  proposed,
see series of papers by G. Sierra  e.g. \cite{Sierra2011, Sierra2012}.

In August 1998, during a conference in Seattle devoted to 100--th  anniversary of the PNT,   Peter Sarnak
offered a bottle of  good  wine for physicists who will be  able to recover some  information from the  Montgomery - Odlyzko
conjecture that is not formerly known to mathematicians. Just two years later he had to go to the store to buy  promised wine.
At the  conference in Vienna in September 1998, Jon Keating delivered a lecture during which  he  announced  solution (but no proof) of
the  so called   problem of  moments of zeta. These results were published later in a joint work with his PhD student
Nina Snaith   \cite{Keating-Snaith}.  For nearly a hundred years mathematicians have tried to calculate moments of the
zeta function on the  critical line
\bee
\frac{1}{T} \int_0^T |\zeta(\tfrac{1}{2}+\imath t)|^{2k}dt, ~~~~~~\text{for  } T\rightarrow \infty
\eee
G.H. Hardy and J.E. Littlewood  \cite[Theorem. 2.41]{H-L} calculated the second moment:
\bee
\frac{1}{T} \int_0^T |\zeta(\tfrac{1}{2}+\imath t)|^{2}dt \sim  \log(T), ~~~~~\text{for  } T\rightarrow \infty.
\label{moment-2}
\eee
The fourth moment  calculated A.E. Ingham  in 1926  \cite[Th. B]{Ingham-1928}
\bee
\frac{1}{T} \int_0^T |\zeta(\tfrac{1}{2}+\imath t)|^{4}dt \sim \tfrac{1}{2\pi} \log^4(T),  ~~~~~\text{for  } T\rightarrow \infty.    
\label{moment-4}
\eee
Higher moments, despite many efforts, were not known, but it was  supposed for $k = 3$   \cite{Conrey-Ghosh-1998}  that:
\bee
 \int_1^T |\zeta(\tfrac{1}{2}+it)|^6~dt \sim \frac{42}{9!} \prod_p\left\{\left(1-\frac1p\right)^{4}\left(1+\frac4p+\frac1{p^2}\right)\right\}T\log^9T \text{~~   for large } T,
\label{moment-6}
\eee
and even more complex expression for  $k=4$ \cite{Conrey+2001}.
\bee
    \int_0^T |\zeta(1/2+it)|^8~dt \sim \frac{24024}{16!}\prod_p\left(\left(1-\frac 1p\right)^{9} \left(1+\frac9p+\frac9{p^2}+\frac1{p^3}\right)\right)T\log^{16}T.
\label{moment-8}
\eee
Keating and Snaith proved the general theorem for  moments of random matrices, which
eigenvalues   have GUE  distribution  and if the behavior of $\zeta(s)$ is modeled by the determinant of such a matrix,
then  their  result applied to the  zeta gives
\bee
\frac{1}{T}\int_0^T \left|\zeta(\tfrac{1}{2}+\imath t)\right|^{2k}\; dt \sim f_k a(k) (\log T)^{k^2},
\eee
where
\bee
a(k) = \prod_{p} \left(1-\frac{1}{p}\right)^{k^2} \sum_{m=0}^\infty \left(\frac{\Gamma(m+k)}{m!\,\Gamma(k)}\right)^2p^{-m} ,
\eee
and numbers $f_k$ are given by
\[
f_k=\frac{G^2(k+1)}{G(2k+1)}.
\]
In the above formula $G(\cdot) $ is the Barnes function satisfying the recurrence  $G(z + 1) = \Gamma(z) G(z) $
with starting value $G(1) = 1$,  thus for natural
arguments this function is a ``factorial over factorials'': $G(n) = 1! \cdot 2! \cdot 3! \ldots (n-2)! $. Of course, the
result of Keating and  Snaith gives formulas \eqref{moment-2}--\eqref{moment-8},  respectively  for $k=1,2, 3, 4$.

In \cite{Crehan-1995} P. Crehan  has shown that for any
sequence of energy levels   obeying a certain growth law  ($|E_n|<e^{an+b}$, for some $a\in \mathbb{R}^+$, $b\in \mathbb{R}$),
there are infinitely many {\it classically  integrable}   Hamiltonians for
which the corresponding quantum spectrum coincides with this sequence. Because from PNT it follows, that the $n-$th prime
$p_n$ grows like $p_n\sim n \log(n)$ the results of Crehan's paper can be applied and there exist classically integrable
hamiltonians whose spectrum coincides with prime numbers, see also \cite{Rosu-2003}.  From \eqref{von-Mangoldt} it follows  that
the imaginary part of the  $n$--th  zero of $\zeta(s)$  grows  like $\gamma_n\sim 2\pi n/\log(n)$,  thus the theorem of Crehan
can be applied and it follows that there  exists an infinite family of classically  integrable  nonlinear oscillators
whose quantum spectrum is given by the imaginary part of the sequence of zeros on the
critical line of the Riemann zeta function.

In the end of XX centaury there were a lot of rumors that Allain Connes has proved the RH using developed by him noncommutative geometry.
Connes \cite{Connes-1999} constructed a quantum system  that has energy levels corresponding to all the Riemann zeros that lie
on the critical line.   To prove RH it has to be shown that  there are no zeros outside critical line, i.e.  unaccounted for
by his energy levels. The operator he constructed acts on a very sophisticated  geometrical space called the noncommutative space of
adele classes.  His approach is very complicated and in fact zeros of the zeta are missing lines (absorption lines) in the
continuous  spectra. During the  passed  time excitement around  Connes's work has faded and much of the hope that his ideas
might lead to the proof of RH  has evaporated.  The common  opinion now is that he
has shifted the problem of  proving the RH to equally difficult problem  of the validity of a certain trace formula.

We mention also the paper  written by S. Okubo\cite{Okubo}  entitled ``Lorentz-Invariant Hamiltonian and Riemann
Hypothesis''.  It is not exactly the realization of the idea of Polya and Hilbert:  the appearing in this paper
two dimensional  differential operator (hamiltonian $H$) does not possess as eigenvalues imaginary parts of the nontrivial
zeros of the $\zeta(s)$. Instead  the  special  condition for zeros of zeta function is used as the boundary
condition for solutions of the eigenvalue equation $H |\phi\rangle=\lambda |\phi\rangle$. Unfortunately, the obtained
eigenfunctions are not normalizable.

Let us remark that for trivial zeros $-2n$ of $\zeta(s)$  with a constant  gap 2 between them  it is possible to
construct  hamiltonian  reproducing these zeros  as eigenvalues.  Namely, the eigenvalue  problem for the harmonic oscillator
in  the units  $m=\hbar=\omega=1$  has the form:
\bee
\frac{1}{2}\left(-\frac{d^{\,2}}{dx^2}+x^2\right)\psi_n(x)=(n+\tfrac{1}{2})\psi_n(x)
\label{oscylator}
\eee
where
\[
\psi_n(x)=e^{ - \tfrac{x^2}{2}} \cdot H_n(x), \qquad n = 0,1,2,\ldots
\]
and $H_n(x)$ are Hermite polynomials:
\[
H_n(x)=(-1)^n e^{x^2}\frac{d^n}{dx^n}\left(e^{-x^2}\right).
\]
Multiplying \eqref{oscylator} by  $-2$  and rearranging terms we obtain
\[
\left(\frac{d^{\,2}}{dx^2}-x^2+1\right)\psi_n(x)=-2n\psi_n(x).
\]
Hence the hamiltoniam for  trivial  zeros of  $\zeta(s)$  is
\[
\hat{H}_{triv}=\frac{d^{\,2}}{dx^2}-x^2+1.
\]

Since the advent of quantum computers and the discovery by Peter Shor of the quantum algorithm  for integer factorization \cite{Shor1995}
there is an interest in applying these algorithms to diverse of problems.  Is it possible to devise the quantum  computer
verifying the RH? We mean here something more clever than, say,  simply mixing the Shor's algorithm
with Lagarias criterion.   Recently there appeared the paper  \cite{Ramos-2013}, in  which authors (assuming the RH) have
built  an unitary  operator with eigenvalues equal to  combination  of nontrivial zeros $\overline{\rho_j}/\rho_j$  lying on the
unit  circle.   Next the quantum circuit representing  this unitary matrix is constructed.
Recently in \cite[p. 4]{Latorre-2015}  the quantum  computer verifying RH was proposed, but it seems to us to be artificial
and not sufficiently  sophisticated: it is based on the \eqref{Koch} and it counts in a quantum way  actual number of prime
numbers   below $x$ and looks for departures  beyond the bound in \eqref{Koch}.

We do not have space here to discuss the use of $\zeta(s)$ in the theory of  Casimir  effect --- devoted to this subject is
the extensive  review by K.  Kirsten in  \cite{Kirsten-2010}, or in  string theory \cite{Cacciatori-2010, Angelantonj-strings}.

\section{Statistical Mechanics and RH} 
\label{knauf}

The partition function   $Z(\beta)$  is the basic quantity used in statistical physics, here  $\beta=1/k_B T$:
$k_B=1.3806488\ldots \times 10^{-23}$[J/K]  is the   Boltzmann constant   and   $T$ is the absolute   temperature.
All thermodynamical  functions can be expressed as derivatives of $Z(\beta)$. The phase transitions appear at such
temperatures that   $Z(\beta)=0$.   For the system,  which may be in micro--states  with energy $E_n$  and can exchange  heat
with environment and with  fixed number of particles, volume and temperature, the  partition function  is  given by
the  formula:
\bee
Z(\beta)=\sum_n e^{-\beta  E_n}.
\label{suma-stanow}
\eee
It turns out that for certain systems  $Z(\beta) $ satisfies  the relation similar to functional  equation  for $\zeta (s)$
and  positions of zeros of the  partition function  analytically continued to  the  whole  complex plane are  highly restricted,
for example to the circle.   These two facts have become the starting point for attempts to prove HR.

It is very easy to construct the system with the $\zeta(s)$ as a partition function. The problem of construction of a simple
one--dimensional Hamiltonian whose spectrum coincides with the set of primes was considered in \cite{Mussardo},
\cite{Sekatskii-2007}, \cite{Schumayer-et-al},  see  also  review  \cite{Rosu-2003}. Some modification should lead to the
Hamiltonian $ H$ having eigenstates $|p\rangle$  labeled by the prime numbers $p$ with eigenvalues $E_p=\mathcal{E}\log(p)$, where
$\mathcal{E}$ is some constant with dimension of energy.  The $n$ particle state can be decomposed into the states $|p\rangle$
using the factorization theorem \eqref{fta}.   The energy  of the state $|n\rangle$ is  equal to
$E(n)=\mathcal{E}\sum_{i=1}^{k}\alpha_i  \log(p_i)=\mathcal{E}\log(n)$. Then the  partition function $Z$ is given by
the Riemann zeta function:
\[
    Z(T) = \sum_{n=1}^\infty \exp \left(\frac{-E_n}{k_B T}\right) = \sum_{n=1}^\infty \exp \left(\frac{-\mathcal{E}\log n}{k_B T}\right) = \sum_{n=1}^\infty \frac{1}{n^s} = \zeta (s), ~~~~s\equiv \mathcal{E}/k_BT
\]
Such a gas was considered  e.g. in  \cite{Spector-1990} and \cite{Bakas-et-al} and found applications
in the string theory.

The functional equation \eqref{functional-zeta} can be written in non--symmetrical form:
\[
2 \Gamma (s)  \cos\big( \frac{\pi}{2} s\big)\zeta (s) = (2\pi)^s \zeta (1-s).
\]
In this form it is  analogous to the Kramers--Wannier \cite{Kramers} duality relation
for the  partition function $Z(J)$  of the two dimensional Ising model with parameter
$J$ expressed in units of $k_B T$ (i.e. equal to interaction constant  multiplied by   $\beta=1/k_B T$)
\bee
Z(J)=2^N (\cosh(J))^{2N} (\tanh(J))^{N}Z(\widetilde{J}),
\eee
where $N$ denotes the number of spins and $\widetilde{J}$ is related to $J$ via
$e^{-2\widetilde{J}} = \tanh(J)$, see e.g. \cite{Feynman-1972}. On the other hand there are so called ``Circle theorems'' on the
zeros of partition functions of some particular systems. To pursue this  analogy  one has to express the
partition function  by the $\zeta(s)$ function.  Then one can hope to prove RH by invoking the
Lee--Yang circle theorem on the zeros  of the partition function.  The Lee--Yang theorem concerns the phase transitions
of some spin systems in external magnetic field and some other models (for a review see \cite{Bena-Droz}).
Let $Z(\beta, z)$ denote the grand -- canonical  partition function, where  $z=e^{\beta H}$ is the fugacity connected with
the magnetic field $H$.
Phase transitions are connected with the singularities of the derivatives of $Z(\beta, z)$, and they appear when $Z(\beta, z)$
is zero. The finite sum defining $Z(\beta, z)$           
can not be a zero for real $\beta$ or $z$ and the Lee--Yang theorem  \cite{Lee-Yang-1952a, Lee-Yang-1952b} asserts that in the
thermodynamical limit, when  the sum for partition function  involves infinite number of terms, all zeros of $Z(\beta, z)$ for
a class of spin models are imaginary and  lie in the complex plane of  the magnetic field $z$ on the unit circle:  $|z|=1$.
The study of zeros of the canonical ensemble  in the complex plane of temperature $\beta$  was initiated by M.Fisher \cite{Fisher-1965}.
He  found  in the thermodynamic limit for a special Ising model not immersed in the magnetic field,  that the zeros of the
canonical partition function also lie on an unit circles, this time in the plane of the complex variable $v = \sinh(2J\beta)$,
where $J > 0$ is the  ferromagnetic coupling  constant.  The critical line $s=\frac{1}{2}+it$ can be mapped into the unit circle
via the transformation $s \rightarrow u=s/(1-s)=(\frac{1}{2}+it)/(\frac{1}{2}-it)$  because then $|u|=1$.   Thus by devising
appropriate spin system with $Z(\beta, z)$ expressed by  the $\zeta(s)$ the Lee--Yang theorem can be used to locate  the possible
zeros of the latter function  and lead to the proof of RH.

In the series of papers A. Knauf  \cite{Knauf-1993, Knauf-1994, Knauf-1998} has undertaken the above outlined plan to attack
the RH. In these papers he introduced the spin system with the partition function in the thermodynamical limit expressed
by zeta function: $Z(s)=\zeta(s-1)/\zeta(s)$ with  $s$ interpreted as the inverse of temperature. However the form of interaction
between  spins  in his model does not belong to one of the cases for which the circle theorem was proved.  This idea was further
developed  in paper \cite{Fiala}.  The authors  of the paper \cite{Planat-2011}  have   shown that RH is  equivalent to an
inequality satisfied by the Kubo–-Martin–-Schwinger states of the Bost and Connes  quantum statistical dynamical system in
special range of temperatures.  There are many other appearances of the $\zeta(s)$ in the statistics of
bosons and fermions, theory of the Bose--Einstein condensate, some special ``number theoretical''   gases etc,
for introduction see \cite[chap. III E]{Physics-and-RH-RevModPhys}.

\section{Random walks,  billiards, experiments etc.}
\label{RW}

The M{\"o}bius function defined in \eqref{mobius} takes  only three values: -1, 0 and 1.  The  values $\mu(n)=1$ and $\mu(n)=-1$
are equiprobable with  probabilities $3/\pi^2\approx 0.3039$,  thus the probability of value  $\mu(n)=0$ is
$1-6/\pi^2\approx 0.3921$. Using values 1 and -1 of the M{\"o}bius function instead of heads or tails of a coin should hence
generate  a symmetric one-dimensional random walk. The total displacement during $n$ steps of such a   random walk will be given
by the summatory function of the M{\"o}bius function: $M(x)=\sum_{n<x} \mu(n)$, which is called the Mertens function.
It is well known that the  ``root mean square''  distance from the starting point of the symmetrical random walk during $N$
steps  grows like   $\sqrt{N}$. The resemblance of $M(n)$ to the symmetrical random walk led F. Mertens in the end of XIX century to
make the conjecture that $M(n)$ grows not faster than the mean displacement of the symmetrical random walk, i.e. $|M(n)|<\sqrt{n}$.
It is an easy calculation to show that  Mertens conjecture implies the RH (vide \eqref{inverse-zeta}):
\[
\frac{1}{\zeta(s)}=\sum_{n=1}^\infty \frac{\mu(n)}{n^s}=\sum_{n=1}^\infty \frac{M(n)-M(n-1)}{n^s}
=\sum_{n=1}^\infty  M(n)\left(\frac{1}{n^s} -\frac{1}{(n+1)^s}\right)=
\]
\[
=\sum_{n=1}^\infty  M(n) \int_n^{n+1}\frac{s dx}{x^{s+1}}=s \sum_{n=1}^\infty \int_n^{n+1}\frac{M(x)dx}{x^{s+1}}
=s \int_1^\infty \frac{M(x)dx}{x^{s+1}}.
\]
If $ M(x)<\sqrt{x}$  then the last integral above gives $ \left|\frac{1}{\zeta(s)}\right|<\frac{|s|}{\sigma-\frac{1}{2}}$,
thus to the right of the line $\Re[s]=\frac{1}{2}$ the inverse of zeta function is bounded hence there can not be zeros
of $\zeta(s)$  in this region and the truth of RH  follows.  For many years mathematicians hoped to prove the RH by showing
the validity of  $|M(n)|<\sqrt{n}$. However in 1985 A. Odlyzko and H. te Riele  \cite{odlyzko1985}   disproved  the Mertens
conjecture; in the proof they have used values of first 2000  zeros of $\zeta(s)$  calculated with accuracy 100--105 digits;
these calculations took  40 hours on  CDC CYBER 750 and  10 hours on Cray-1  supercomputers. Using Mathematica these
computations can be done on the modern laptop in a couple of minutes.  Littlewood proved that the RH is equivalent to
slightly modified Mertens conjecture
\[
M(n)= {\mathcal{O}}(n^{1/2+\epsilon}) \Leftrightarrow {\rm RH~ is ~true}.
\]
The fact that $M(n)$ behaves like a one dimensional
random walk was also pointed out in \cite{Churchhouse1968} and used  to show that RH is ``true with probability 1''.

In the paper \cite{Shlesinger} M. Shlesinger has investigated a very special one-dimensional random walk which can be linked
with the RH. The probability of jumping to other sites with  steps having a displacement of $\pm l$ sites involves   directly
the M{\"o}bius function:
\[
p(\pm {\rm \textit{l}}) = \frac{1}{2} C \left(\frac{1}{\rm {\textit{l}}^{~1+\beta}}
 \pm \frac{\mu(\rm {\textit{l}})}{\rm {\textit{l}}^{~1+\beta-\epsilon}}\right)   , ~~~~\beta>0,
\]
where $C=\frac{1}{\zeta(1+\beta)+\frac{1}{\zeta(1+\beta)}}$ is a normalization factor, $\beta$ (to be not  confused
with $\beta=1/k_B T$)  is the fractal dimension of the set of points visited by random  walker.
He coined the name Riemann--M{\"o}bius  for this random walk.  Some general properties of
the "structure function" $\lambda(k)$ being the Fourier of the probabilities $p(l)$:   $\lambda(k)=\sum_l e^{ikl}p(l)$,
enabled Shlesinger to locate the complex zeros inside the critical strip, however the result of  J. Hadamard
and  Ch. J. de la Vallée--Poussin that  $\zeta(1 + it)\ne 0$ can not be recovered by this method. What is interesting the
existence of off critical line zeros is not in  contradiction with behavior of $\lambda(k)$ following from the universal
laws of probability.

In \cite{Alexander-et-al}  the stochastic interpretation of the Riemann zeta function was given.  There are much more
connections between $\zeta(s)$ and random walks as well as Brownian motions known to mathematicians. The
extensive review of obtained results  expressing expectation values of different random variables by $\zeta(s)$ or $\xi(s)$ can
be found in \cite{Biane-et-al}.


In the paper \cite{Bunimovich}  L.A. Bunimovich  and  C.P. Dettmann considered  the  point particle
bouncing inside the circular billiard. There is a possibility that the small  ball will escape through a small hole on the
reflecting perimeter. Let $P_1(t)$ denote the  probability  of not escaping  from a circular
billiard with one hole till time $t$.  Bunimovich  and Dettmann obtained exact  formula for $P_1(t)$ and surprisingly
this probability was expressed by $\zeta(s)$.  So here again the function of purely number theoretical origin meets the physical
reality. Then they proved that  RH is equivalent  to
\bee
\lim_{\epsilon\to 0}\lim_{t\to\infty} \epsilon^{\delta}(tP_1(t)-2/\epsilon)=0
\eee
be true for every  $\delta>-1/2$.  Here this value $1/2$ is directly connected with the location of  critical line in the
formulation of RH.   A little bit more complicated condition was obtained for biliard with two holes.
In principle such conditions  allow experimental verification of RH using microwave cavities simulating  billiards or optical
billiards  constructed with  microlasers. Experiments can refute RH if the behavior of   $tP_1(t)-2/\epsilon$
in  the limit $\epsilon\to 0$ will be slower than power like dependence $\epsilon^{1/2}$  in the limit of vanishing $\epsilon$.
To our knowledge up today no such experiments were performed. In the paper \cite{Dettmann-2014}  generalization to the
spherical billiard was considered. Again the survival probability in such a 3D biliard is related to the Riemann hypothesis.

\begin{wrapfigure}{l}[0pt]{0pt}
\includegraphics[width=0.4\textwidth, angle=0]{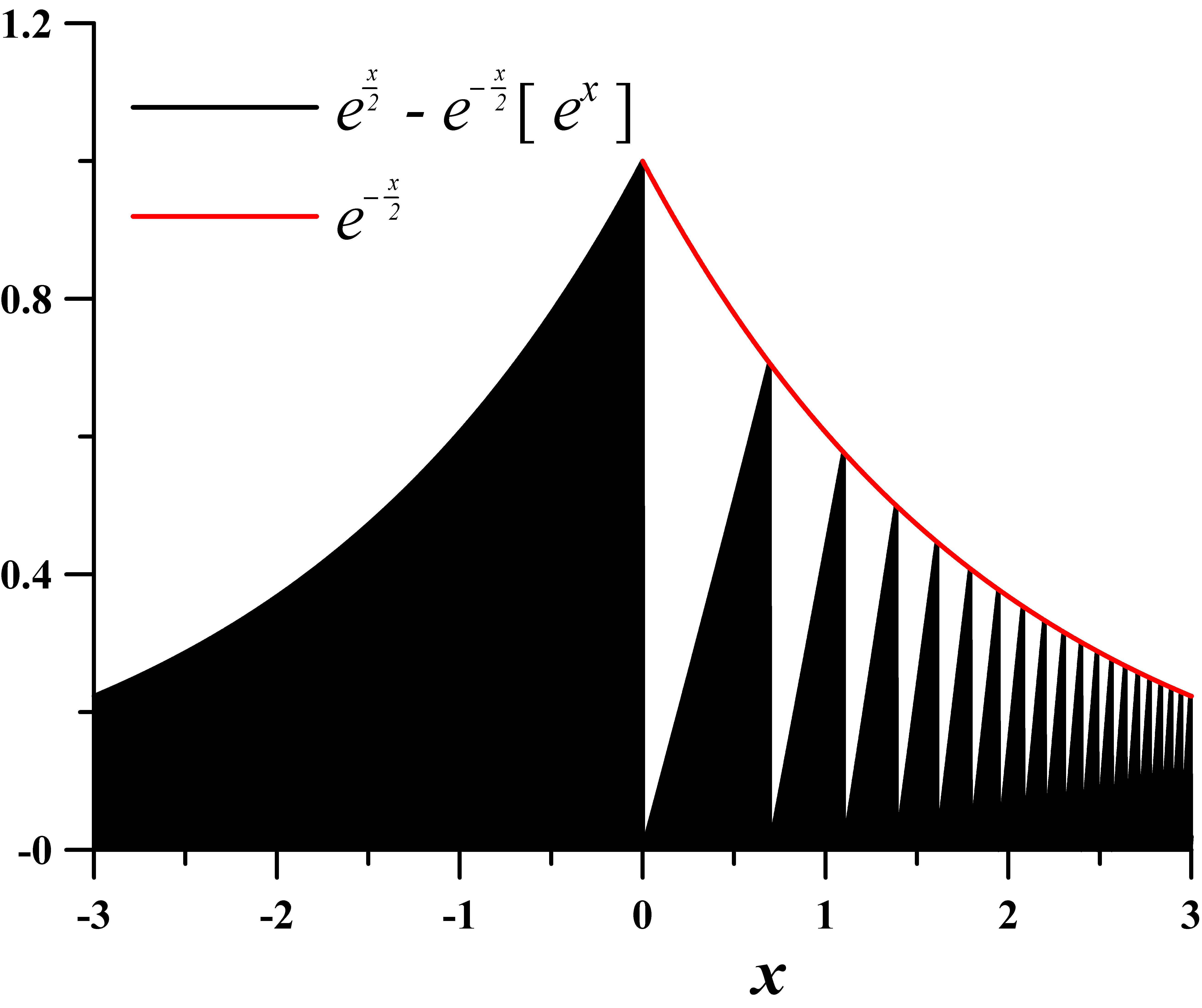} \\
\caption{\small Plot of the function appearing in the integral  representation \eqref{van der Pol} of the zeta function.}
\label{machine}
\baselineskip 28pt
\end{wrapfigure}

Already in 1947  van der Pol has built the electro--mechanical device verifying the RH \cite{VanderPol}. He has
built machine  plotting the $\zeta(1/2+it)$ from the following integral representation:
\bee
\frac{\zeta(\frac{1}{2}+it)}{\frac{1}{2}+it}= \int_{-\infty}^{\infty}\left( e^{-x/2}\lfloor e^x \rfloor - e^{x/2}\right) e^{-ixt} dx.
\label{van der Pol}
\eee
Here $\lfloor x \rfloor$   denotes  integer part  of $x$.  It has the form of Fourier transform of the function
$y(x)=e^{-x/2}\lfloor e^x \rfloor - e^{x/2}$.  The plot of integrand is
shown in Fig. \ref{machine}.  The shape of this function was cut precisely with scissors on the edge of a paper disc. The beam of
light was passing between teeth on the perimeter of the disc and detected by the photocell. The resulting from photoelectric effect
current was superimposed with current of varying frequency to perform analogue Fourier transform. After some additional operations
van der Pol has obtained the plot of modulus  $| \zeta(\frac{1}{2}+it)/\frac{1}{2}+it|$ on which  the first
29  nontrivial zeta  zeros  were located with accuracy better than \%1.  The authors of \cite{Physics-and-RH-RevModPhys}
have summarized this experiment in the words: ``This construction, despite its limited achievement, deserves to be treated
as a gem in the history of the natural sciences.''.

It is well known that two--dimensional electrostatic fields can be found using the functions of complex variables. There arises
a question to which electrostatic problem the zeta function can be linked? In the recent paper \cite{Leclair2013} A. LeClair
has developed this analogy and he constructed a two--dimensional vector field $\overrightarrow{E}$ from the real and
imaginary parts of the zeta function. It allowed him to derive the formula for the $n$-th zero on the critical line of $\zeta(s)$
for large $n$  expressed as the solution of a simple transcendental equation.

In the  written \cite{Dyson_frogs}  version of his AMS Einstein Lecture ``Birds and frogs'' (which was to have been given in
October 2008 but which unfortunately had to be canceled) Freeman Dyson points to the possibility of proving the RH using
the similarity  in  behavior  between one--dimensional  quasi-crystals  and  the  zeros  of the $\zeta(s)$  function.
If RH is true then locations of its nontrivial zeros would define a one--dimensional quasi--crystal but  the classification
of them is still missing.

\section{Zeta is a fractal}
\label{fraktal}

In 1975 S.M. Voronin  \cite{Voronin-theorem}  proved remarkable theorem on the universality of the Riemann $\zeta(s)$ function.

{\bf Voronin's theorem}:  Let  $0 < r < 1/4$ and  $f(s)$  be a  complex function continuous for  $|s| \leq r$ and
analytical in the interior of the disk.  If   $f(s) \neq 0$, then for every   $\epsilon > 0$ there exists real number
$T=T(\epsilon,f)$  such that:
\bee
\max_{|s| \leq r} \; \left| f(s) - \zeta\left(s+\left(\frac{3}{4} + i\;T\right)\right)
 \right| < \epsilon.
\eee

\begin{wrapfigure}{l}[0pt]{0pt}
\includegraphics[ width=0.47\textwidth, angle=0]{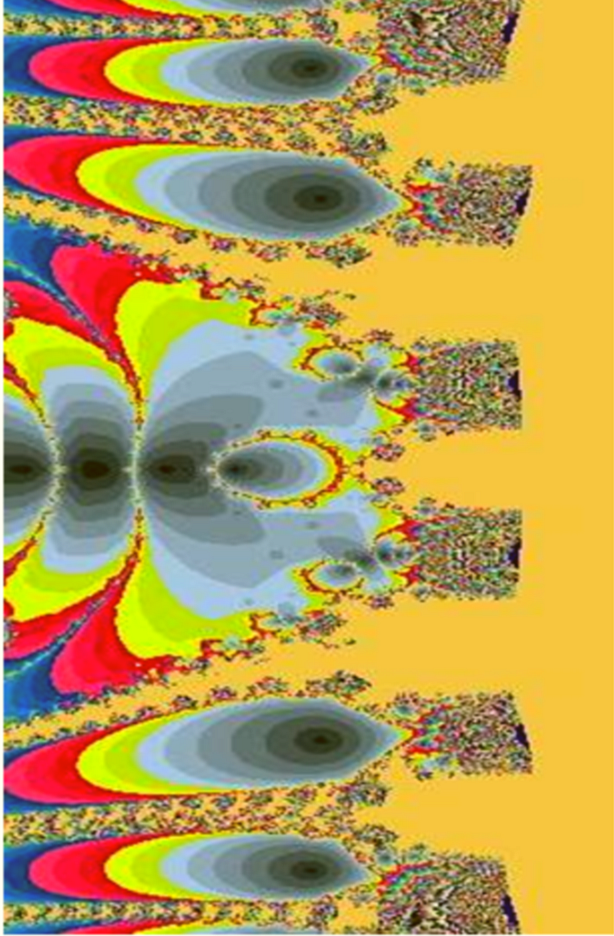}    
\caption{\small The plot of initial values for \eqref{Kawahira} $-9<\Re(z_0)<9, -25<\Im(z_0)<25$ showing in colors the number
$n$  of iterations of \eqref{Kawahira}  after which $|\zeta(z_n)|< 0.000001$. In black are shown regions around zeros of zeta. Clearly are visible
zeros:  -2, -4, -6, -8 and nontrivial $\frac{1}{2}\pm i14.134\ldots, \frac{1}{2}\pm i21.022\ldots $.}
\label{Dukiewicz}
\end{wrapfigure}

Put simply in words it means that the zeta function approximates  locally any smooth function in a uniform way!  By applying
this theorem to itself,  i.e.  taking as $f(s)=\zeta(s)$, we obtain that $\zeta(s)$ is selfsimilar, see  S.C. Woon  \cite{Woon-fraktal}
who has shown that
{\it the Riemann $\zeta(s)$ is a fractal.}  In the paper \cite{Bitar-1991}  the Voronin's theorem was applied to the physical
problem:  to propose a new formulation  of the Feynman's path integral.

Another aspect of fractality  of zeta was found in \cite{Wu-Sprung-1993, Schumayer-et-al},  where the one--dimensional quantum
potential was numerically constructed from known zeta zeros which in turn are reproduced  as eigenvalues of  this potential.
The fractal dimension of the graph of this potential was determined to be around 1.5. In \cite{Schumayer-et-al}  even
the multifractal nature of this potential was revealed.

In the late  seventies of XX centaury John Hubbard has analysed the Newton's method for  finding approximations to the roots
of equation $f(x)=0$ to the case of polynomial $z^3-1$ on the complex plane. In this method the root $x^\star$ of $f(x^\star)=0$
is obtained as a limit $x^\star = \lim_{n \rightarrow \infty}x_n$ of the sequence:
\[
    x_{n+1} = x_n - \frac{f(x_n)}{f'(x_n)}.
\]

If the function $f(x)$ has a few roots the limit depends on the choice of the initial $x_0$.  Hubbard was interested in
the question which starting points $z_0 \in \mathbb{C}$ tend to one of three roots $1, (-1\pm \sqrt{3}i)/2$ of $z^3=1$.  He obtained
one of the first fractal images full of interwoven corals. Tomoki  Kawahira has applied Newton's method to the Riemann's
zeta function:
\bee
 z_{n+1} = z_n - \frac{\zeta(z_n)}{\zeta'(z_n)}.
\label{Kawahira}
\eee
Because $\zeta(s)$ has infinitely many roots, instead of looking for basin domains of different zeta zeros, he looked for the number
of iterations of  \eqref{Kawahira}  for a given starting point $z_0$ needed to fall into the close vicinity of one of  the zeros.
Let  us mention that such a  modification was also applied to the original problem $z^3=1$. He obtained beautiful pictures
representing the zeros of $\zeta(s)$. We present  in Fig. \ref{Dukiewicz} the plot obtained by M. Dukiewicz \cite{Dukiewicz}.

\section{Concluding Remarks}

We have given many examples of physical problems connected to the RH.  In XIX centaury all these problems were not known,
but it seems that  Riemann  believed that the questions of mathematics could be answered with the help of  physics
and in fact he performed  some  physical experiments by himself  to check some of his theorems,  see \cite{Elizalde2013}.
We add here, that there is a wide spread  rumor among the people who are trying to solve the RH that
Fields Medal Laureate Enrico Bombieri  believes that RH will be proved by a physicist, see \cite[p. 4]{Sautoy2003}.

Some mathematicians enunciate the opinion that RH is not true because long open conjectures in analysis tend to be false.
In other words nobody has proved RH because simply it is not true.  There are examples from number theory when
some conjectures confirmed by huge  ``experimental'' data finally  turned out to be false  and possible counterexamples are so
large that never will be accessible to computers.  One such common belief was the inequality ${\Li}(x)>\pi(x)$ remarked
already by Gauss and confirmed by all available data,  now it is about $x=10^{18}$. However, in 1914 J.E. Littlewood has
shown \cite{Littlewood} that the difference between the number of primes smaller than $x$ and the
logarithmic integral up to $x$ changes the sign infinitely many times, what was another rather complicated proof of the infinitude
of primes.  The smallest value $x_S$ such that  for the first time $\pi(x_S)\geq {\Li}(x_S)$ holds is called Skewes number
because in 1933 S. Skewes \cite{SkewesI},  assuming the truth of the Riemann hypothesis,  argued that it is certain that
$\pi(x)-{\Li}(x)$ changes sign for some  $x_S < 10^{10^{10^{34}}}$. In 1955 Skewes \cite{SkewesII} has found, without assuming
the Riemann hypotheses, that $\pi(x)-{\Li}(x)$ changes sign at some $x_S < \exp\exp\exp\exp(7.705)<10^{10^{10^{10^3}}}$.
This enormous bound for $x_S$ was several times lowered and the lowest present day known estimation of the Skewes number
is around $10^{316}$, see \cite{Bays} and \cite{Demichel2}.
The second example is  provided by the Mertens conjecture discussed in Sect.\ref{RW}. The inequality $|M(x)|<x^\frac{1}{2}$
is confirmed by all available data but finally it is false. Like in the case of the inequality $\pi(x)>{\Li}(x)$
we can expect first $x$ for which  $|M(x)|>x^\frac{1}{2}$  at horribly heights. Namely  J. Pintz \cite{Pintz1987}  has shown that
the first counterexample appears below $\exp(3.21 \times 10^{64})$. This upper bound was later
lowered to $\exp(1.59\times 10^{40})$ \cite{KotnikRiele2006}.   Such examples show that confirmation of some facts up to say
$10^{18}$ is misleading and  somewhere at $t=10^{10^{\iddots}}$ the nontrivial  zero of $\zeta(s)$  with  real part
different from $\frac{1}{2}$  can be lurking.

\bigskip

\noindent{\bf Acknowledgement}: I thank Magdalena Dukiewicz--Jurka for allowing me  to use the Fig. \ref{Dukiewicz} from her
master thesis (performed under my guidance in 2006).  I thank  Magdalena Za{\l}uska--Kotur and Jerzy  Cis{\l}o
for comments and remarks.


\end{document}